\documentclass[12pt,oneside]{book}
\usepackage{amsmath}
\usepackage{amsthm}
\usepackage{graphics}
\usepackage{epsfig,amssymb,latexsym,verbatim}
\usepackage{graphicx}
\usepackage{multirow}
\usepackage{multicol}
\usepackage{float}
\usepackage{booktabs}
\usepackage{fancyhdr}
\usepackage{enumerate,verbatim}
\usepackage[sort&compress]{natbib}
\usepackage{rotating}
\usepackage{hyperref}
\usepackage{subfigure}
\usepackage{ulem}
\usepackage{url}
\usepackage{threeparttable}
\usepackage{xr}
\usepackage{multirow}
\usepackage{multicol}
\usepackage{wrapfig}
\usepackage{lscape}
\usepackage{rotating}
\usepackage{epstopdf}
\usepackage{subfigure}
\usepackage{amsfonts}
\usepackage[usenames,dvipsnames]{color}
\usepackage{times,color}
\usepackage{amsbsy}
\usepackage{amsthm}
\usepackage[sf,small,center, compact]{titlesec}
\usepackage[lined,algonl]{algorithm2e}

\newcommand{\cc}{{\tiny\rm c}}

\def\bs{\boldsymbol}

\definecolor{red}{rgb}{1,0,0}
\definecolor{green}{rgb}{0,1,0}
\definecolor{blue}{rgb}{0,0,1}

\definecolor{lxy}{RGB}{180,0,180}

\definecolor{gw}{RGB}{0,76,153}

\newcommand{\D}{{\tiny\rm D}}
\newcommand{\I}{{\tiny\rm I}}
\newcommand{\R}{{\tiny\rm R}}

\floatstyle{ruled}
\newfloat{algorithm}{tbp}{loa}
\floatname{algorithm}{Algorithm}
\def\bs{\boldsymbol}

\newcommand{\thetheorem}{{\thesection. \arabic{theorem}}}
\newcommand{\thelemma}{{\thesection. \arabic{lemma}}}
\newcommand{\theproposition}{{\thesection. \arabic{proposition}}}
\newcommand{\thecorollary}{{\thesection. \arabic{corollary}}}

\begin{document}
\renewcommand{\baselinestretch}{1.25}
\markboth{\hfill{\footnotesize\rm L. Wang, G. Wang, L. Gao, X. Li, S. Yu, M. Kim, Y. Wang and Z. Gu}\hfill}
{\hfill {\footnotesize\rm } \hfill}
\renewcommand{\thefootnote}{}

\fontsize{10.95}{14pt plus.8pt minus .6pt}\selectfont
\vspace{0.8pc} \centerline{\Large\bf Spatiotemporal Dynamics, Nowcasting and Forecasting}
\centerline{\Large\bf  of COVID-19 in the United States} \vspace{.4cm}
\centerline{
Li Wang$^{a}$, Guannan Wang$^{b}$, Lei Gao$^{a}$, Xinyi Li$^{c}$, Shan Yu$^{d}$, Myungjin Kim$^{a}$,} 
\centerline{Yueying Wang$^{a}$ and Zhiling Gu$^{a}$}
\vspace{.4cm}
\centerline{\it $^{a}$Iowa State University, USA, $^{b}$College of William \& Mary, USA, $^{c}$Clemson University, USA}
\centerline{\it and $^{d}$University of Virginia, USA}
\vspace{%
.55cm} \fontsize{9}{11.5pt plus.8pt minus .6pt}\selectfont
\footnote{\emph{Address for correspondence}: Li Wang (lilywang@iastate.edu) 
}

\begin{quotation}
\noindent \textit{Abstract:} Epidemic modeling is an essential tool to understand the spread of the novel coronavirus and ultimately assist in disease prevention, policymaking, and resource allocation. In this article, we establish a state of the art interface between classic mathematical and statistical models and propose a novel space-time epidemic modeling framework to study the spatial-temporal pattern in the spread of infectious disease. We propose a quasi-likelihood approach via the penalized spline approximation and alternatively reweighted least-squares technique to estimate the model. Furthermore, we provide a short-term and long-term county-level prediction of the infected/death count for the US by accounting for the control measures, health service resources, and other local features. Utilizing spatiotemporal analysis, our proposed model enhances the dynamics of the epidemiological mechanism and dissects the spatiotemporal structure of the spreading disease. To assess the uncertainty associated with the prediction, we develop a projection band based on the envelope of the bootstrap forecast paths. The performance of the proposed method is evaluated by a simulation study. We apply the proposed method to model and forecast the spread of COVID-19 at both county and state levels in the United States.

\vspace{9pt} \noindent \textit{Key words and phrases:} Coronavirus;
nonparametric modeling;
partially linear models;
spatial epidemiology;
splines;
varying coefficient models.
\end{quotation}

\fontsize{10.95}{14pt plus.8pt minus .6pt}\selectfont

\thispagestyle{empty}

\setcounter{chapter}{1} 
\setcounter{section}{0} 
\renewcommand{\thesection}{\arabic{section}} %
\renewcommand{\thesubsection}{1.\arabic{subsection}}%
\setcounter{equation}{0} \renewcommand{\theequation}{1.\arabic{equation}} %
\setcounter{table}{0} \renewcommand{\thetable}{{1.\arabic{table}}} %
\setcounter{figure}{0} \renewcommand{\thefigure}{1.\arabic{figure}} %
\setcounter{algorithm}{0} \renewcommand{\thealgorithm}{1.\arabic{algorithm}} %
\setcounter{theorem}{0} \renewcommand{\thetheorem}{{1.\arabic{theorem}}} %
\setcounter{lemma}{0} \renewcommand{\thelemma}{{1.\arabic{lemma}}} %
\setcounter{proposition}{0} \renewcommand{\theproposition}{{1.\arabic{proposition}}}%
\setcounter{corollary}{0} \renewcommand{\thecorollary}{{1.\arabic{corollary}}}%

\section{Background \& Introduction}
The severe acute respiratory syndrome coronavirus 2 (SARS-CoV-2) outbreak started last December, and it has expanded to impact nearly every corner of the world. On  New Year's Eve of 2019, the World Health Organization (WHO) was informed of mysterious pneumonia cases in Wuhan, China. On January 3, 2020, 44 cases were reported to WHO, among whom 11 were severely ill \citep{WHO_unknown:20}. By February 4, confirmed cases had been reported in 24 countries outside China \citep{WHO_remarkFeb:20}. The US confirmed its first case in Washington state On January 21 -- a man who returned to the US from Wuhan \citep{cdc_firstUScase:20}. On January 30, the US Centers for Disease Control and Prevention (CDC) confirmed the person-to-person spread of coronavirus in the US \citep{cdc_p2p:20}. On the same day, the WHO declared the coronavirus outbreak as a Public Health Emergency of International Concern \citep{who_emer:20,who_timeline:20}.

Many nonpharmaceutical interventions (NPIs) were implemented to prevent the spread of COVID-19. For example, Wuhan implemented screening measures for travelers leaving the city at airports, railway stations, and other passenger terminals, and eventually closed off Wuhan City on January 22 \citep{nyt_wuhan_lockDown:20}. The US started screening at twenty airports at the end of January \citep{wp_airport:20}. On February 25, San Francisco became the first US city to declare a state of emergency over COVID-19 \citep{sf_gov:20}, followed by the states of Washington \citep{wa_gov:20} and Florida \citep{fl_gov:20}. On March 13, President Donald Trump declared a national COVID-19 emergency \citep{national_emergency:20}, and sixteen states announced school closures by then \citep{Ujifusa2020}. On March 19, California issued a ``stay-at-home'' order for all of its 40 million residents \citep{KarimiMoon2020}, and within two weeks, the majority of the states had taken similar actions. By the end of March, more than 91\% of the world's population lived in countries with restrictions for non-resident travelers from abroad \citep{pew:20}.

Even with all the control measures taken in place, the spread of COVID-19 is still dramatic. From February 7 to 14, both the total cases confirmed and deaths worldwide almost doubled within one week. Meanwhile, COVID-19 kept spreading globally, and over 50 countries reported confirmed cases by the end of February \citep{WHO_febReport:20}. During mid-March, COVID-19 presented in all 50 states in the US Starting from March 26, the US led the world in COVID-19 cases. On May 28, the US COVID-19 death count passed one hundred thousand, and then in the middle of June, the number of confirmed cases of COVID-19 hit two million in the US. The confirmed and death cases kept increasing rapidly in the following months. By the beginning of September, the US surpassed six million confirmed cases and hit seven million on September 25. On October 16, the US surpassed eight million confirmed cases and 218 thousand deaths \citep{cdc_dashboard:20}.

The effect of COVID-19 is profound. The World Bank estimated that the coronavirus pandemic could push additional 16 million people into extreme poverty. On April 14, the International Monetary Fund warned that the world is facing its worst economic downturn as coronavirus lockdowns continue to wreak havoc on the global economy \citep{imf:20}. Due to the immense pressures of the crippled economy and an anxious public amid a pandemic, the US started to loosen lockdown measures in late April and backtracked after reopening for a few weeks and seeing a surge in cases. As the pandemic progresses, there are broader interests from the public in answering questions such as how far the SARS-CoV-2 virus will spread, how many lives it will eventually claim, how effective intervention strategies will be, as well as whether and when the pandemic will resurge.

Epidemic modeling is an important scientific tool to answer these questions by aiding people to understand the pandemic data, make predictions, and help the medical professionals and decision-makers allocate resources and design/evaluate intervention strategies to fight against COVID-19; see \cite{Gog:20, Vespignani:etal:20}. Several attempts had been made to model and forecast the spread and mortality of COVID-19; for example, see  \cite{Kucharski:etal:20,Sun:etal:20}.

The fundamental concept of infectious disease epidemiology is investigating how the diseases spread. Mathematical models are undeniably useful in understanding the dynamics of infectious disease spread \cite{Keeling:Rohani:08}. An essential type of mathematical model is the class of mechanistic models such as the  Susceptible-Infectious-Removed (SIR) compartmental model or the Susceptible-Exposed-Infectious-Recovered model (SEIR); see details in \cite{Brauer:etal:08, Lawson:etal:16}. Mechanistic models make explicit hypotheses about the biological mechanisms that drive the dynamics of infection. These mechanistic models characterize disease transmission through a set of differential equations, and they demonstrate the average behavior of the epidemic. Statistical modeling is another powerful tool for extracting information about the disease spread in epidemic studies \cite{Held:etal:20}. The data modeling is one of the cultures in statistical modeling, and it is usually designed for inference about the relationships between variables whilst also catering to prediction. When analyzing the confirmed cases and deaths of COVID-19, other factors, such as demographics, socioeconomic status, mobility, and control policies, may also be responsible for temporal or spatial patterns. For maximal effectiveness, we create a novel method to appreciate and exploit the complementary strengths of mathematical and statistical models in this paper. 

We borrow the mechanistic rules from the SIR model and form a data-driven model with three compartments: infectious, susceptible, and removed states. The capacity of the health care system, and control measures, such as government-mandated social distancing, also have a significant impact on the spread of the epidemic. We borrow the strength from statistical models and include various explanatory variables to study not only the spatiotemporal structure but also the effects of the explanatory variables. Notice that the spread of the disease varies a lot across different geographical regions; we incorporate discrete-time spatially varying coefficient models to different compartments to reconstruct the spatiotemporal dynamics of the disease transmission. In general, the spatiotemporal models are able to bring in more information to the epidemic study \citep{held2019handbook, Jia:etal:2020}.

With an emerging disease such as COVID-19, it is hard to measure many features of the transmission process, which may take a long time to fully understand. Thus, it is desirable to make inferences from observed data as model-free as possible. For a parametric epidemic model, the typical inference problem involves estimating the parameters associated with the parametric models from the data at hand. Such specifications are ad hoc, and if misspecified, can lead to substantial estimation bias problems. This issue might be addressed in practice by considering alternative nonparametric models or sensitivity analyses if some of the underlying model parameters are assumed to be known. By adopting a nonparametric approach, we do not impose a particular parametric structure, which significantly enhances the flexibility of the parametric epidemic models. Nonparametric approaches to fitting epidemic models to the data have received relatively little attention in the literature, possibly due to the lack of data. With the rich COVID-19 epidemic data released every day, we can consider the nonparametric method to model the covariates and coefficient functions.

By allowing the response (such as infected and death counts) to depend on time and location, we consider a generalized additive varying coefficient model to estimate the unobserved process of the disease transmission. For our model estimation, we propose a quasi-likelihood approach via the penalized spline approximation and an iteratively reweighted least-squares technique. Our proposed algorithm is sufficiently fast and efficient for the user to analyze large datasets within seconds. As an empirical illustration, we apply the proposed model and estimation method to a study of COVID-19 at the county-level in the US. We illustrate how the proposed method can be used to analyze the spatiotemporal dynamics of the disease spread and guide evidence-based decision making.

Finally, prediction models for COVID-19 at the county-level that combine local characteristics and actions are very beneficial for the community to understand the dynamics of the disease spread and support decision making at a time when they are urgently needed. Knowing more about the vulnerable communities and the reasons for those communities that are more likely to be infected are crucial for the policy and decision-makers to assist in prevention efforts and ultimately stop the pandemic. In this paper, we consider both the short-term impact and long-term impact of the SARS-CoV-2 virus at the county level in the US. To assess the uncertainty associated with the prediction, we also propose a projection band constructed based on the envelope of the bootstrap forecast paths, which are closest to the forecast path obtained based on the original sample.

The rest of the paper is organized as follows. Section \ref{sec:data} presents the epidemic and endemic data. Section \ref{sec:model} outlines the nonparametric spatiotemporal modeling framework and describes how to incorporate additional covariates. Section \ref{sec:estimation} introduces our estimation method, our algorithm, and details of the implementation. Section \ref{sec:prediction} provides the prediction as well as the uncertain quantification with the prediction band of the forecast path. Section \ref{sec:simulation} evaluates the finite sample performance of the proposed method using a simulation study. Section \ref{sec:application} describes the results and findings of the case study. Section \ref{sec:discussion} concludes the paper with a discussion. Supplementary Material (\url{https://faculty.sites.iastate.edu/lilywang/arxiv}) contains some animation videos of the dynamic estimation results of the COVID-19 study.

\setcounter{chapter}{2} 
\setcounter{section}{1} 
\renewcommand{\thesection}{\arabic{section}} %
\renewcommand{\thesubsection}{2.\arabic{subsection}}%
\setcounter{equation}{0} \renewcommand{\theequation}{2.\arabic{equation}} %
\setcounter{table}{0} \renewcommand{\thetable}{{2.\arabic{table}}} %
\setcounter{figure}{0} \renewcommand{\thefigure}{2.\arabic{figure}} %
\setcounter{algorithm}{0} \renewcommand{\thealgorithm}{2.\arabic{algorithm}} %
\setcounter{theorem}{0} \renewcommand{\thetheorem}{{2.\arabic{theorem}}} %
\setcounter{lemma}{0} \renewcommand{\thelemma}{{2.\arabic{lemma}}} %
\setcounter{proposition}{0} \renewcommand{\theproposition}{{2.\arabic{proposition}}}%
\setcounter{corollary}{0} \renewcommand{\thecorollary}{{2.\arabic{corollary}}}%

\section{COVID-19 Case Study and Data} \label{sec:data}

\subsection{Research Goal of the Study}

The goal of this study is threefold. First, we develop a new dynamic epidemic modeling framework for public health surveillance data to study the spatial-temporal pattern in the spread of COVID-19. We aim to investigate whether the proposed model could be used to guide the modeling of the dynamic of the spread at the county level by moving beyond the typical theoretical conceptualization of context where a county's infection is only associated with its own features. Second, to understand the factors that contribute to the spread of COVID-19, we model the daily infected cases at the county level in consideration of the demographic, environmental, behavioral, socioeconomic factors in the US. Third, we project the spatial-temporal pattern of the spread of the virus in the US. For the short-term forecast, we provide the prediction of the daily infection count and death count up to the county level. As for the long-term forecast, we project the total infected and death cases in the next three months.

\subsection{Epidemic Data from the COVID-19 Outbreak in the US} 

This study analyzes data from the reported confirmed COVID-19 infections and deaths at the county level, which are reported by the 3,104 counties from the 48 mainland US states and the District of Columbia. The aggregated COVID-19 cases are from January 20 until September 3, 2020. The data are collected, compiled and cleaned from a combination of public sources that aim to facilitate the research effort to confront COVID-19, including Health Department Website in each state or region, the New York Times \citep{NYT:20}, the COVID-19 Data Repository by the Center for Systems Science and Engineering at Johns Hopkins University \citep{JHUCSSE}, and the COVID Tracking Project \citep{COVIDTrack}. These data sources automatically updated every day or every other day. We have created a dashboard \url{https://covid19.stat.iastate.edu/} to visualize and track the infected and death cases, which was launched on March 27, 2020.

\subsection{Information of the covariates}  \label{ssec:covariates}

We consider various county-level characteristics as covariate information in our study, which can be divided into six groups. The data sources and the operational definitions of these features are discussed as follows, and a list of all the variables is summarized in Table \ref{Tab:covariates}.

\textbf{Policies}.
Government declarations are used to identify the dates that different jurisdictions implemented various social distancing policies (emergency declarations, school closures, bans on large gatherings, limits on bars, restaurants and other public places, the deployment of severe travel restrictions, and ``stay-at-home'' or ``shelter-in-place'' orders). President Trump declared a state of emergency on March 13, 2020, to enhance the federal government response to confront the COVID-19. By March 16, 2020, every state had made an emergency declaration. Since then, more severe social distancing actions had been taken by the majority of the states, especially those hardest hit by the pandemic. We compiled the dates of executive orders by checking national and state governmental websites, news articles and press releases. 

\textbf{Demographic Characteristics}. To capture the demographic characteristics of a county, five variables are considered in the analysis to describe the racial, ethnic, sexual and age structures: the percent of the population who identify as African American, the percent of the population who identify as Hispanic or Latino, the rate of aged people ($\geq 65$ years) per capita, the ratio of male over female and population density over square mile of land area. The former two variables were obtained from the 2010 Census (US Census Bureau, 2010), and the latter three variables are extracted from the 2010--2018 American Community Survey (ACS) Demographic and Housing Estimates (US Census Bureau, 2010).

\textbf{Healthcare Infrastructure}. We incorporated three components in our analysis to describe the healthcare infrastructure in each county: percent of the population aged less than 65 years without health insurance, local government expenditures for health per capita, and total counts of hospital beds per 1,000 population. These components measure the access for residents to public health resources within and across counties. The first component is available in the USA Counties Database (US Census Bureau, 2010), the second is from Economic Census 2012 (US Census Bureau, 2012), and the last is compiled from Homeland Infrastructure Foundation-level Data (US Department of Homeland Security).

\textbf{Socioeconomic Status}. A diverse of factors are considered to describe the socioeconomic status in each county. We first apply the factor analysis to seven factors collected from the 2005--2009 ACS 5-year estimates (US Census Bureau, 2010), and generate two factors: social affluence and concentrated disadvantage. To be specific, the former is comprised of the percent of families with annual incomes higher than \$75,000 (factor loading = 0.86), percent of the population aged 25 years or older with a bachelor's degree or higher (factor loading = 0.92), percent of the people working in management, professional, and related occupations (factor loading = 0.73), and the median value of owner-occupied housing units (factor loading = 0.74); whereas the latter includes the percent of the households with public assistance income (factor loading = 0.34), the percent of households with female householders and no husband present (factor loading = 0.81), and civilian labor force unemployment rate (factor loading = 0.56). These two factors, affluence and disadvantage, explain more than 60\% of the variation. 

We also incorporate the Gini coefficient to measure income inequality. The Gini coefficient, also known as Gini index, is a well-known measure for income inequality and wealth distribution in economics, with value ranging from zero (complete equality, where everyone has exactly the same income) to one (total inequality, where one person occupies all of the income). The 2005--2009 ACS (US Census Bureau, 2010) provided the household income data that allow us to calculate the Gini coefficient.

\textbf{Rural/urban Factor}.
In the literature, rural/urban residence has been found to be associated with the spread of epidemics. Specifically, rural counties are often characterized by poor socio-economic profiles and limited access to healthcare services, indicating a potential higher risk. To capture rural/urban residence, we use the urban rate from the 2010 Census (US Census Bureau, 2010).

\textbf{Geographic Information}. 
The longitude and latitude of the geographic center for each county in the US are available in Gazetteer Files (US Census Bureau, 2019).  

\textbf{Mobility}. The data are collected and cleaned from the US Department of Transportation,  Bureau of Transportation Statistics and Descartes Labs. It describes the daily number of trips within each county produced from an anonymized national panel of mobile device data from multiple sources. Trips are defined as movements that include a stay of longer than 10 minutes at an anonymized location away from home. 

\renewcommand{\baselinestretch}{1.25}
\begin{table}[htbp]
\caption{County-level characteristics (predictors) used in the modeling.}
\centering
\begin{tabular}{p{2.5cm}p{12.5cm}}
	\hline 
	Predictors & Description \\
	\hline
	Control & Dummy variable for declaration of ``shelter-in-place" or ``stay-at-home'' order (Control$~=1$, for ``shelter-in-place'', and  Control$~=0$, for no restriction or restriction lifted) \\ \hline
	\multicolumn{2}{l}{Socioeconomic Status}\\
	Affluence & Social affluence \\
	Disadvantage & Concentrated disadvantage \\
	Gini & Gini coefficient
	\\ \hline
	\multicolumn{2}{l}{Healthcare Infrastructure}\\
	NHIC & Percent of persons under 65 years without health insurance\\
	EHPC & Local government expenditures for health per capita\\ 
	TBed$^{\ast}$ & Total bed counts per 1000 population \\ \hline
	\multicolumn{2}{l}{Demographic Characteristics}\\
	AA & Percent of African American population \\
	HL & Percent of Hispanic or Latino population \\
	PD$^{\ast}$ & Population density per square mile of land area \\
	Old & Aged people (age $\geq 65$ years) rate per capita \\
	Sex & Ratio of male over female \\ \hline
	\multicolumn{2}{l}{Environment Characteristics}\\
	Mobility &  Daily number of trips within each county \\
	Urban & Urban rate\\  
	\hline 
\end{tabular}
\label{Tab:covariates}
\begin{tablenotes}
\small 
\item Note: The covariates with $^{\ast}$ represent that they are transformed from the original value by $f(x)=\log (x+\delta)$. For example, $\text{PD}^{\ast}=\log(\text{PD}+\delta)$, where $\delta$ is a small number.
\end{tablenotes}
\end{table}

\setcounter{chapter}{3} 
\setcounter{section}{2} 
\renewcommand{\thesection}{\arabic{section}} %
\renewcommand{\thesubsection}{3.\arabic{subsection}}%
\setcounter{equation}{0} \renewcommand{\theequation}{3.\arabic{equation}} %
\setcounter{table}{0} \renewcommand{\thetable}{{3.\arabic{table}}} %
\setcounter{figure}{0} \renewcommand{\thefigure}{3.\arabic{figure}} %
\setcounter{algorithm}{0} \renewcommand{\thealgorithm}{3.\arabic{algorithm}} %
\setcounter{theorem}{0} \renewcommand{\thetheorem}{{3.\arabic{theorem}}} %
\setcounter{lemma}{0} \renewcommand{\thelemma}{{3.\arabic{lemma}}} %
\setcounter{proposition}{0} \renewcommand{\theproposition}{{3.\arabic{proposition}}}%
\setcounter{corollary}{0} \renewcommand{\thecorollary}{{3.\arabic{corollary}}}%

\section{Space-time Epidemic Modeling} \label{sec:model}

In this section, to study the spatiotemporal pattern of COVID-19, we developed a novel spatiotemporal epidemic model (STEM) to estimate and predict the infection and death cases at the area level based on the idea of the compartment models. For simplicity, we introduce the STEM based on the parsimonious SIR models, but it can be extended to the SEIR models with an extra ``exposed" compartment for infected but not infectious individuals.

For area $i$ and day $t$, let $Y_{it}$ be the number of new cases, and let $I_{it}$, $D_{it}$, $R_{it}$, and $S_{it}$ be the number of accumulated active infectious cases, accumulated death cases, accumulated recovered cases, and susceptible population, respectively. Let $N_{i}$ be the total population for the $i$th area, and denote $Z_{it}=\log(S_{i t}/N_{i})$. Let $\mathbf{U}_{i}=(U_{i1},U_{i2})^{\top}$ be the GPS coordinates of the geographic center of area $i$, which ranges over a bounded domain $\Omega \subseteq \mathbb{R}^2$ of the region under study. Let $\mathbf{X}_{i}=(X_{i1}, \ldots,X_{iq})^{\top}$ be a $q$-dimensional vector of explanatory variables collected from the US Census Bureau. For example, the socioeconomic factors, health resources, and demographic conditions. Let $A_{ijt}$ be the $j$th dummy variable of actions or measures taken for area $i$ at time $t$, and let $\mathbf{A}_{it}=(A_{i1t},\ldots,A_{ipt})^{\top}$, which varies with the time. 

In this paper, we consider the exponential families of distributions. The conditional density of $Y_{it}$ given $(I_{i-1},Z_{i,t-1},\mathbf{A}_{i,t-r},\mathbf{X}_{i},\mathbf{U}_{i}) =(w, z, \bs{a},\bs{x},\bs{u})$ can be represented as 
\begin{equation*}
f\left(\left. y\right| w,z,\bs{a},\bs{x},\bs{u}\right) =\exp \left[\frac{1}{\sigma^2}\left\{y\zeta \left(w, z,\bs{a},\bs{x},\bs{u}\right) -%
\mathcal{B}\left\{\zeta \left(w, z,\bs{a},\bs{x},\bs{u}\right)\right\} \right\}+%
\mathcal{C}\left(y, \sigma^2 \right) \right],
\end{equation*}
for some known functions $\mathcal{B}$ and $\mathcal{C}$, dispersion parameter $\sigma^2$ and the canonical parameter $\zeta$.  

We assume that the determinants of the daily new cases of a particular area can be explained not only by the features of that area but also by the characteristics of the surrounding areas. Based on the idea of the SIR models, we propose a discrete-time spatial epidemic model comprising the susceptible state, infected state, removed state, and area-level characteristics. Below we use superscripts $\I$, $\D$ and $\R$ to denote infected, death and recovered states. We assume that the conditional mean value of daily new positive cases ($\mu_{it}^{\I}$), fatal cases ($\mu_{it}^{\D}$) and recovery ($\mu_{it}^{\R}$) in area $i$ and day $t$ can be modeled via a link function $g$ as follows: 
\begin{align}
    g(\mu_{it}^{\I}) &= \beta_{0t}^{\I} (\mathbf{U}_i) + \beta_{1t}^{\I} (\mathbf{U}_i) \log(I_{i,t-1}) + \alpha_{0t}^{\I} Z_{i,t-1} +\sum_{j=1}^p\alpha_{jt}^{\I} A_{ij,t-r}+\sum_{k=1}^q \gamma_{kt}^{\I}(X_{ik}), \label{model:STEM-infection}\\
    g(\mu_{it}^{\D}) &=\beta_{0t}^{\D} (\mathbf{U}_i) + \beta_{1t}^{\D}\log(I_{i,t-\delta})+\sum_{j=1}^p\alpha_{jt}^{\D}A_{ij,t-r^{\prime}}+\sum_{k=1}^q \gamma_{kt}^{\D}(X_{ik}), \label{model:STEM-death} \\
    \mu_{it}^{\R} &= \nu_{t}^{\R} I_{i, t-\delta^{\prime}},
    \label{model:STEM-recover}
\end{align}
where $\alpha_{jt}^{\I}$'s, $\alpha_{jt}^{\D}$'s, $\beta_{1t}^{\D}$ and $\nu_{t}^{\R}$ are unknown constant coefficients, $\beta_{0t}^{\I}(\cdot)$, $\beta_{1t}^{\I}(\cdot)$, and $\beta_{0t}^{\D}(\cdot)$ are unknown bivariate coefficient functions, $\gamma_{kt}^{\I}(\cdot)$, $\gamma_{kt}^{\D}(\cdot)$, $k=1,\ldots,q$, are univariate functions to be estimated, $\delta$ and $\delta^{\prime}$ are the time delay between illness and death or recovery, and the parameter $r$ in $A_{ij,t-r}$'s denotes a small delay time allowing for the control measure to be effective (here we take $r=7$, $r^{\prime}=7$). For model identifiability, we assume $\mathrm{E}(\gamma_{kt}^{\I})=0$, $\mathrm{E}(\gamma_{kt}^{\D})=0$, $k=1,\ldots, q$. The STEM encompasses many existing models as special cases. For example, the traditional generalized linear regression models, generalized additive models \citep{Liu:Yang:Hardle:13},  generalized partially linear additive models \citep{Wang:Liu:Liang:Carroll:11} and generalized additive coefficient model \citep{xue2010polynomial}.

Note that, for the log link,  $\exp\{\beta_{0t}^{\I}(\bs{u})\}$ illustrates the transmission rate at location $\bs{u}$, $\exp\{\beta_{0t}^{\D}(\bs{u})\}$ represents the fatality rate at location $\bs{u}$, $\nu_{t}^{\R}$ is the recovery rate, $\beta_{1t}^{\I}(\cdot)$, $\alpha_{0t}^{\I}$ and $\beta_{1t}^{\D}$ are the mixing parameters of the contact process. The rationale for including $\beta_{1t}^{\I}(\cdot)$ and $\beta_{1t}^{\D}$ ($\beta_{1t}^{\I}(\cdot)>0$, $\beta_{1t}^{\D}>0$) is to allow for deviations from mass action and to account for the discrete-time approximation to the continuous time model \citep{Finkenstadt:etal:00,Wakefield:19}. In many cases, the standard bilinear form may not necessarily hold. The above proposed epidemic model incorporates the nonlinear incidence rates, which represents a much wider range of dynamical behavior than those with bilinear incidence rates \citep{Liu:Hethcote:Levin:87}. These dynamical behaviors are determined mainly by $\beta_{0t}^{\I}(\cdot)$, $\beta_{1t}^{\I}(\cdot)$, $\beta_{0t}^{\D}(\cdot)$, and $\beta_{1t}^{\D}$. For example, when $\beta_{1t}^{\I}(\cdot)$ and $\alpha_{0t}^{\I}$ are both $1$, it corresponds to the standard assumption of homogeneous mixing in \cite{De:etal:95}. 

In our study, since we model the number of new cases at time $t$ for area $i$, Poisson or Negative Binomial (NB) might be an appropriate option for random component \citep{Yu:etal:20,Kim:Wang:20}. For example, for the infection model, we can assume that 
\begin{itemize}
\item (Poisson) $\mathrm{E}(Y_{it} | Z_{i,t-1}, \mathbf{A}_{i,t-r},\mathbf{X}_i, \mathbf{U}_{i})= \mu_{it}^{\I}$, $\mathrm{Var}(Y_{it} | Z_{i,t-1}, \mathbf{A}_{i,t-r},\mathbf{X}_i, \mathbf{U}_{i}) = \mu_{it}^{\I}$,
\item (NB) $\mathrm{E}(Y_{it} | Z_{i,t-1}, \mathbf{A}_{i,t-r},\mathbf{X}_i, \mathbf{U}_{i})= \mu_{it}^{\I}$, $\mathrm{Var}(Y_{it} | Z_{i,t-1}, \mathbf{A}_{i,t-r},\mathbf{X}_i, \mathbf{U}_{i}) = \mu_{it}^{\I}(1 + \mu_{it}^{\I}/I_{i,t-1})$, 
\end{itemize}
where $\mu_{it}^{\I}$ can be modeled via the same log link as follows:
\begin{equation}
\label{model:STEM_infection}
\log(\mu_{it}^{\I})= \beta_{0t}^{\I} (\mathbf{U}_{i}) + \beta_{1t}^{\I} (\mathbf{U}_{i}) \log(I_{i,t-1}) + \alpha_{0t}^{\I} Z_{i,t-1} +\sum_{j=1}^p\alpha_{jt}^{\I} A_{ij,t-r}+\sum_{k=1}^q \gamma_{kt}^{\I}(X_{ik}).
\end{equation} 
We can consider similar models for the death count. At the beginning of the outbreak, infected and death cases could be rare,  so ``Poisson" might be a reasonable choice of the random component to describe the distribution of rare events in a large population. As the disease progresses, the variation of infected/death count increases across counties and states. So, at the acceleration phase of the disease, the negative binomial random component might be an appropriate option for the presence of over-dispersion.

The above spatiotemporal epidemic model (STEM) is developed based on the foundation of epidemic modeling. It can provide a rich characterization of different types of errors for modeling the uncertainty. Moreover, it accounts for both spatiotemporal nonstationarity and area-level local features simultaneously. It also offers flexibility in assessing the dynamics of the spread at different times and locations than various parametric models in the literature.

\setcounter{chapter}{4} 
\setcounter{section}{3} 
\renewcommand{\thesection}{\arabic{section}} %
\renewcommand{\thesubsection}{4.\arabic{subsection}}%
\setcounter{equation}{0} \renewcommand{\theequation}{4.\arabic{equation}} %
\setcounter{table}{0} \renewcommand{\thetable}{{4.\arabic{table}}} %
\setcounter{figure}{0} \renewcommand{\thefigure}{4.\arabic{figure}} %
\setcounter{algorithm}{0} \renewcommand{\thealgorithm}{4.\arabic{algorithm}} %
\setcounter{theorem}{0} \renewcommand{\thetheorem}{{4.\arabic{theorem}}} %
\setcounter{lemma}{0} \renewcommand{\thelemma}{{4.\arabic{lemma}}} %
\setcounter{proposition}{0} \renewcommand{\theproposition}{{4.\arabic{proposition}}}%
\setcounter{corollary}{0} \renewcommand{\thecorollary}{{4.\arabic{corollary}}}%

\section{Estimation of the STEM} \label{sec:estimation}

In this section, we describe how to estimate the parameters and nonparameteric components in the proposed STEM model. To capture the temporal dynamics, we consider the moving window approach. 

\subsection{Penalized Quasi-likelihood Method}

We first describe the estimation of model (\ref{model:STEM-infection}). For the current time $t$, and rounghness parameters $\lambda_{0}$ and $\lambda_{1}$, we consider the penalized quasi-likelihood problem defined as follows:
\begin{align}
	\sum_{i=1}^{n}\sum_{s=t-t_0}^{t}&L\left[g^{-1}\left\{\beta_{0} (\mathbf{U}_{i}) + \beta_{1} (\mathbf{U}_{i})\log(I_{i,s-1}) + \alpha_{0} Z_{i,s-1}+\sum_{j=1}^p\alpha_{j}A_{ij,s-r} \right. \right.\nonumber \\
	&\quad\quad\quad\quad\quad\quad\quad\quad\quad\left.\left. +\sum_{k=1}^q \gamma_{k}(X_{ik}) \right\}, Y_{is}\right] -\frac{1}{2}\left\{\lambda_{0}\mathcal{E}(\beta_{0})+\lambda_{1}\mathcal{E}(\beta_{1})
\right\},
\label{DEF:quasilikelihood}
\end{align}
where $t_0+1$ is the window width for the model fitting, and it can can be selected by minimizing the prediction errors or maximizing the correlation between the predicted and observed values. The energy functional is defined as follows: 
\begin{equation}
	\mathcal{E}(\beta)
	=\int_{\Omega} \left\{(\nabla_{u_{1}}^{2}\beta)^2+2(\nabla_{u_{1}}\nabla_{u_{2}}\beta)^2+(\nabla_{u_{2}}^{2}\beta)^2\right\}\mathrm{d}u_{1}\mathrm{d}u_{2},
\label{EQ:energyfun}
\end{equation}
where $\nabla_{u_{j}}^{q}\beta(\bs{u})$ is the $q$th order derivative in the direction $u_{j}$, $j=1,2$, at any location $\bs{u}=(u_1,u_2)^{\top}$.

Note that, except for parameters $\{\alpha_{j}\}_{j=0}^{p}$, other functions are related to curse of dimensionality due to the nature of functions. To handle this difficulty, we employ the basis expansion approach to approximate the univariate and bivariate functions discussed below. The univariate additive components $\{\gamma_{k}(\cdot)\} _{k=1}^{q}$ and the spatially varying coefficient components $\{\beta_{\ell}(\cdot) \}_{\ell=0}^{1}$ in model (\ref{model:STEM_infection}) are approximated using univariate polynomial spline and bivariate penalized splines over triangulation (BPST), respectively. The BPST method is well known to be computationally efficient to deal with data distributed on complex domains with irregular shape or with holes inside; see the details in \cite{Lai:Wang:13} and \cite{Sangalli:Ramsay:Ramsay:13}. We provide a brief introduction to univariate and bivariate splines below. Assume that $X_k$ takes value on an interval $[a_k,b_k]$, $k=1,\ldots, q$. 
Let $\mathcal{U}_{k}=\mathcal{U}_k^{\varrho}([a_k,b_k])$ be the space of the polynomial splines of order $\varrho+1$. Next, let $\mathcal{U}_{k}^{0}=\{\phi \in \mathcal{U}_{k}: \mathrm{E}\phi(X_{k}) = 0\}$, which creates a space of the centered spline functions; see \cite{xue2010polynomial}, \cite{Liu:Yang:Hardle:13} and \cite{Wang:Xue:Yang2020}. Let $\mathcal{J}$ be the index set of the basis functions, and then denote by $\{\varphi_{kJ}(x_k),J\in \mathcal{J}\}$ the original B-spline basis functions for the $k$th covariate. Let
$\varphi_{kJ}^0(x_k)=\varphi_{kJ}(x_k)-\varphi_{k1}(x_k)\mathrm{E}\varphi_{kJ}(X_k)/\mathrm{E}\varphi_{k1}(X_k)$, $\Phi_{kJ}(x_k)=\varphi_{kJ}^{0}(x_k)/\text{SD}\{\varphi_{kJ}^{0}(X_k)\}$, $J\in \mathcal{J}$, then $\mathrm{E}\Phi_{kJ}(X_k)=0$ and $\mathrm{E}\Phi_{kJ}^{2}(X_k)=1$. For all $x_{k} \in [a_k,b_k]$, the estimator of $\gamma_k(x_{k})$ is $\widehat{\gamma}_k(x_{k}) = \sum_{J\in \mathcal{J}}\xi _{kJ}\Phi_{kJ}(x_{k})=\bs{\Phi}_{k}^{\top}(x_{k}) \bs{\xi}_{k}$, where $\bs{\Phi}_{k}(x_{k})=\left(\Phi_{kJ}(x_{k}),J\in \mathcal{J}\right)^{\top}$ and $\bs{\xi}_{k}=\left(\xi_{kJ}, J\in \mathcal{J}\right)^{\top}$ is a vector of coefficients.

For the bivariate coefficient functions $\beta_{0}\left(\cdot \right)$ and $\beta_{1}\left(\cdot \right)$ in the STEM model (\ref{model:STEM_infection}), we introduce bivariate spline over triangulation \citep{Lai:Wang:13}. The spatial domain $\Omega$ with either an arbitrary shape or holes inside can be partitioned into $M$ triangles, $T_1,\ldots,T_M$, that is, $\Omega=\cup_{m=1}^{M} T_m$. Then, denote $\triangle :=\{T_1,\ldots,T_M\}$ by a triangulation of the domain $\Omega$ \citep{Lai:Schumaker:07}; see, for example, Figure \ref{FIG:triangulations}. 
Given an integer $d\geq 0$, let $\mathbb{P}_{d}(T)$ be the space of all polynomials of degree $\leq d$ on $T$. 
Let $\mathbb{C}^r(\Omega)$ be the space of $r$th continuously differentiable functions over the domain $\Omega$. For $0\leq r < d$ and $\triangle$, we construct the spline space of degree $d$ and smoothness $r$ over $\triangle$ in the following:
\begin{equation}
\mathbb{S}_{d}^{r}(\triangle)=\{\mathcal{P} \in \mathbb{C}^{r}(\Omega):\mathcal{P} |_{T_m}\in \mathbb{P}_{d}(T_m), T_m \in \triangle, m=1,\ldots, M\}.
\label{EQ:smoothspace}
\end{equation}

For triangulation $\triangle$ with $M$ triangles, 
denote a set of bivariate Bernstein basis polynomials for $\mathbb{S}_{d}^{r}(\triangle)$ as $\{B_{m}(\bs{u})\}_{m \in \mathcal{M}}$, where an index set $\mathcal{M}$ for basis functions has its cardinality $|\mathcal{M}|=M(d+1)(d+2)/2$. Then, we can approximate the bivariate functions $\beta_{\ell}(\cdot) \in \mathbb{S}_{d}^{r}(\triangle)$ in the STEM model (\ref{model:STEM_infection}) by $\sum_{m \in \mathcal{M}} B_{m}(\bs{u})\theta_{\ell m}
    =\mathbf{B}(\bs{u})^{\top}\bs{\theta}_{\ell}$, where $\mathbf{B}(\bs{u})=\{B_{m}(\bs{u}), m \in \mathcal{M}\}^{\top}$  and $\bs{\theta}_{\ell} =\left\{\theta_{\ell m}, m \in \mathcal{M} \right\}^{\top}$. 

In practice, the triangulation can be obtained through varieties of software; see for example, the ``Delaunay'' algorithm (\textit{delaunay.m} in MATLAB or \textit{DelaunayTriangulation} in MATHEMATICA), the R package ``Triangulation'' \citep{Triangulation}, and the ``DistMesh'' Matlab code. The bivariate spline basis are generated via the R package ``BPST'' \citep{BPST}. See more discussions of how to generate the bivariate spline over triangulations in \cite{ Mu:Wang:Wang:18} and \cite{Yu:etal:20}. 

Considering the basis expansion, the energy functional  $\mathcal{E}(\beta_{\ell })$ in (\ref{EQ:energyfun}) can be approximated by $\mathcal{E}(\mathbf{B}(\cdot)^{\top}\bs{\theta}_{\ell})=\bs{\theta}_{\ell}^{\top}\mathbf{P}\bs{\theta}_{\ell}$, for $\ell=0,1$, where $\mathbf{P}$ is a block diagonal penalty matrix. By introducing the constraint matrix $\mathbf{H}$ which satisfies $\mathbf{H}\bs{\theta}_{\ell} = \bs{0}$, $\ell=0,1,$  we can reflect global smoothness in $\mathbb{S}_{d}^{r}(\triangle)$ in (\ref{EQ:smoothspace}). For the current time $t$, the maximization problem (\ref{DEF:quasilikelihood}) is changed to minimize
\begin{align}
	-&\sum_{i=1}^{n}\sum_{s=t-t_0}^{t} L\left(g^{-1}\left[\mathbf{B}(\mathbf{U}_{i})^{\top}\{\bs{\theta}_{0} +\bs{\theta}_{1}\log(I_{i, s-1})\} + \alpha_{0} Z_{i,s-1}+\sum_{j=1}^p\alpha_{j}A_{ij,s-r}\right.\right. \label{EQ:const_minimization}
	\\
	&+\left.\left.\sum_{k=1}^q \bs{\Phi}_{k}^{\top}(X_{ik}) \bs{\xi}_{k}\right], Y_{is}\right) +
		\frac{1}{2}(\lambda_{0}\bs{\theta}_{0}^{\top}\mathbf{P}\bs{\theta}_{0}+
		\lambda_{1}\bs{\theta}_{1}^{\top}\mathbf{P}\bs{\theta}_{1}) 
	~~ \text{subject to}~~ \mathbf{H}\bs{\theta}_{\ell} =\bs{0}, \ell=0,1.\nonumber
\end{align}  

Directly solving the optimization problem in (\ref{EQ:const_minimization}) is not straightforward due to the smoothness constraints involved. Instead, suppose that the rank $r$ matrix $\mathbf{H}^{\top}$ is decomposed into $\mathbf{Q}\mathbf{R}=\left(\mathbf{Q}_1 ~\mathbf{Q}_2\right)
{\binom{\mathbf{R}_{1}}{\mathbf{R}_{2}}}$, where $\mathbf{Q}_1$ is the first $r$ columns of an orthogonal matrix $\mathbf{Q}$, and $\mathbf{R}_2$ is a matrix of zeros, which is a submatrix of an upper triangle matrix $\mathbf{R}$. Then, reparametrization of $\bs{\theta}_{\ell} = \mathbf{Q}_2\bs{\theta}^{\ast}_{\ell}$ for some $\bs{\theta}^{\ast}_{\ell}$, $\ell=0,1$, enforces $\mathbf{H}\bs{\theta}_{\ell}=\bs{0}$. Thus, the constraint problem in (\ref{EQ:const_minimization}) can be changed to an unconstrained optimization problem as follows:
\begin{align}
	-\sum_{i=1}^{n}&\sum_{s=t-t_0}^{t}L\left(g^{-1}\left[\mathbf{B}(\mathbf{U}_{i})^{\top}\mathbf{Q}_2\{\bs{\theta}^{\ast}_{0} +\bs{\theta}^{\ast}_{1}\log(I_{i, s-1})\} + \alpha_{0} Z_{i,s-1} +\sum_{j=1}^p\alpha_{j}A_{ij,s-r}\right.\right.\nonumber\\
	&\quad\left.\left.+\sum_{k=1}^q \bs{\Phi}_{k}^{\top}(X_{ik}) \bs{\xi}_{k}\right], Y_{is}\right) 
+\frac{1}{2}\left(\lambda_{0}\bs{\theta}_{0}^{\ast \top}\mathbf{Q}_2^{ \top}\mathbf{P}\mathbf{Q}_2\bs{\theta}^{\ast}_{0}+\lambda_{1}\bs{\theta}_{1}^{\ast \top}\mathbf{Q}_2^{\top}\mathbf{P}\mathbf{Q}_2\bs{\theta}^{\ast}_{1}\right).
	\label{EQ:unconst_minimization}
\end{align}  

Let $(\widehat{\bs{\theta}}^{\ast}_{0 t}, \widehat{\bs{\theta}}^{\ast}_{1 t})^{\top}$, $(\widehat{\alpha}_{0t},\widehat{\alpha}_{1t},\ldots,\widehat{\alpha}_{pt})^{\top}$, and $(\widehat{\bs{\xi}}_{1t},\ldots,\widehat{\bs{\xi}}_{qt})^{\top}$ be the minimizers of (\ref{EQ:unconst_minimization}) at time point $t$. Then, the estimator of $\bs{\beta}_{\ell t}(\cdot)$ is $\widehat{\bs{\beta}}_{\ell t}(\bs{u})=\mathbf{B}(\bs{u})^{\top}\mathbf{Q}_2\widehat{\bs{\theta}}^{\ast}_{\ell t}$, $\ell=0,1$,
the estimator of $\alpha_{jt}$ is
$\widehat{\alpha}_{jt}$, $j=1,\ldots,p$, and the spline estimator $\gamma_{kt}(\cdot)$ is $\widehat{\gamma}_{kt}(x_k)=\bs{\Phi}_{k}(x_k)^{\top}\widehat{\bs{\xi}}_{kt}$, $k=1, \ldots, q$. 

\subsection{Penalized Iteratively Reweighted Least Squares Algorithm} \label{SI:Comp_Alg}

We now describe the estimating algorithm in detail. For the current time $t$, let $\mathbf{Y} = (\mathbf{Y}_{1}^{\top},\ldots , \mathbf{Y}_{t}^{\top})^{\top}$ be the vector of the responses where $\mathbf{Y}_{s} = (Y_{1s},\ldots , Y_{ns})^{\top}$. 
Denote $\bs{\Phi}_i^{\top}=\{\bs{\Phi}_1(X_{i1})^{\top},\cdots, \bs{\Phi}_q(X_{iq})^{\top}\}$, 
$\mathbf{A}_{is}^{\top}=(A_{i1,s-r}, \cdots, A_{ip,s-r})$, 
and $\mathbf{F}=(\mathbf{F}_{1},\ldots,\mathbf{F}_{t})^{\top}$, 
where $\mathbf{F}_{s}=(\mathbf{F}_{1s},\cdots, \mathbf{F}_{ns})$, 
and $\mathbf{F}_{is}^{\top}= (\mathbf{A}^{\top}_{is}, ~\bs{\Phi}_{i}^{\top},~ [\{1,\log(I_{i,s-1})\}^{\top}\otimes\{\mathbf{Q}_2^{\top}\mathbf{B}(\mathbf{U}_{i})\}]^{\top})$. Denote $\bs{\vartheta}=(\bs{\alpha}^{\top}, \bs{\xi}^{\top}, \bs{\theta}^{\ast \top})^{\top}$, and let $\eta_{is}(\bs{\vartheta})
	=\mathbf{B}(\mathbf{U}_{i})^{ \top}\mathbf{Q}_2\{\bs{\theta}_{0}^{\ast} +\bs{\theta}_{1}^{\ast}\log(I_{i,s-1})\}+\alpha_{0}Z_{i,s-1} +\sum_{j=1}^p\alpha_{j}A_{ij,s-r}+\sum_{k=1}^q \bs{\Phi}_{k}^{\top}(X_{ik}) \bs{\xi}_{k}$, 
and $\bs{\eta}(\bs{\vartheta})=\left\{\eta_{is}\right\}_{i=1,s=1}^{n,t}$. 
In addition, let the mean vector 
$\bs{\mu}(\bs{\vartheta})
	=\{\mu_{is}(\bs{\vartheta})\}_{i,s=1}^{n,t}
	=\{g^{-1}\left(\eta_{is}(\bs{\vartheta})\right)\}_{i,s=1}^{n,t}$, 
the variance function matrix $\mathbf{V}=\mathrm{diag}\{V(\mu_{is})\}_{i,s=1}^{n,t}$, the diagonal matrix $\mathbf{G}=\mathrm{diag}\{g^{\prime}(\mu_{is})\}_{i,s=1}^{n,t}$ with the derivative of link function as element, and the weight matrix $\widetilde{\mathbf{V}}=\mathrm{diag}[\{V(\mu_{is})g'(\mu_{is})^2\}^{-1}w_{st},i=1,\ldots,n, s=1,\ldots,t]$, where $w_{st}=I(t-s\geq t_0)$.  
   
In order to efficiently solve the minimization in (\ref{EQ:unconst_minimization}), we design a penalized iteratively reweighted least squares (PIRLS) algorithm as described below. 
Suppose at the $j$th iteration, we have $\bs{\mu}^{(j)} =\bs{\mu}(\bs{\vartheta}^{(j)})$, $\bs{\eta}^{(j)} =\bs{\eta}(\bs{\vartheta}^{(j)})$ and $\mathbf{V}^{(j)}$ . Then at $(j+1)$th iteration, we consider the following objective function:
\[
	L_{P}^{(j+1)}=\left\|\left\{\mathbf{V}^{(j)}\right\}^{-1/2}\left\lbrace \mathbf{Y}-\bs{\mu}\left( \bs{\vartheta}^{(j)}\right) \right\rbrace \right\|^2
	+\frac{1}{2}\sum_{\ell=0}^{1}\lambda_{\ell}\bs{\theta}_{\ell}^{\ast\top}\mathbf{Q}_{2}^{\top}\mathbf{P}\mathbf{Q}_{2}\bs{\theta}_{\ell}^{\ast}.
\]

Take the first order Taylor expansion of $\bs{\mu}(\bs{\vartheta})$ around $(\bs{\vartheta}^{(j)})$, then
\begin{align}
\nonumber
	L_{P}^{(j+1)} &\approx \left\|\left\{\mathbf{V}^{(j)}\right\}^{-1/2}\left[ \mathbf{Y}-\bs{\mu}^{(j)}-\{\mathbf{G}^{(j)}\}^{-1}\mathbf{F}(\bs{\vartheta}- \bs{\vartheta}^{(j)}) \right] \right\|^2+\frac{1}{2}\sum_{\ell=0}^{1}\lambda_{\ell}\bs{\theta}_{\ell}^{\ast\top}\mathbf{Q}_{2}^{\top}\mathbf{P}\mathbf{Q}_{2}\bs{\theta}_{\ell}^{\ast}\\
&= \left\|\left\{\widetilde{\mathbf{V}}^{(j)}\right\}^{1/2} \left[\widetilde{\mathbf{Y}}^{(j)}-\mathbf{F}
(\bs{\vartheta})
\right]\right\|^2
+\frac{1}{2}\sum_{\ell=0}^{1}\lambda_{\ell}\bs{\theta}_{\ell}^{\ast\top}\mathbf{Q}_{2}^{\top}\mathbf{P}\mathbf{Q}_{2}\bs{\theta}_{\ell}^{\ast},
\label{iterativestep}
\end{align}
where $\widetilde{\mathbf{Y}}^{(j)}=(\widetilde{\mathbf{Y}}_{1}^{(j) \top}, \ldots, \widetilde{\mathbf{Y}}_{t}^{(j)\top})^{\top} $ with $\widetilde{Y}_{is}^{(j)}= g^{\prime}(\mu_{is}^{(j)})(Y_{is}-\mu_{is}^{(j)} )+\eta_{is}^{(j)}$ for $s=1,\ldots,t$. The detailed procedure for the PIRLS is illustrated in Algorithm \ref{ALGO:PIRLS}. In the numerical studies, we consider the following initial values   $\mu_{is}^{(0)}=Y_{is}+0.1$ and $\eta_{is}^{(0)}=g(\mu_{is}^{(0)})$ for $i=1,\ldots,n$ and $s=1,\ldots,t$.

\normalem{
\begin{algorithm}
\footnotesize
\SetKwBlock{Begin}{Step 2.}{}
\caption{The Penalized Iteratively Reweighted Least Squares Algorithm.}
\BlankLine
\textbf{Step 1.} Initialize $\bs{\eta}^{(0)}$ and $\bs{\mu}^{(0)}$ and calculate $\widetilde{\mathbf{V}}^{(0)}$ and  $\widetilde{\mathbf{Y}}^{(0)}$ from $g'(\mu_{is}^{(0)})$ and $V(\mu_{is}^{(0)})$, $i=1,\ldots,n,$ and $s=1,\ldots, t$.

\Begin
(Set step $j=0$.)
{
\While {$\left\{\bs{\alpha}, \bs{\xi}, \bs{\theta}^{\ast}\right\}$ \textrm{not converge}}{
~(i) 
Obtain $\bs{\alpha}^{(j+1)}, \bs{\xi}^{(j+1)}, \bs{\theta}^{\ast (j+1)}$ by minimizing the  (\ref{iterativestep}) with respect to $\bs{\vartheta}$, and  $\bs{\eta}^{(j+1)}=\bs{\eta}(\bs{\vartheta}^{(j+1)})$ and $\bs{\mu}^{(j+1)}=\bs{\mu}(\bs{\vartheta}^{(j+1)})$.\\
\noindent (ii) 
Update $\widetilde{\mathbf{V}}^{(j+1)}$ and $\widetilde{\mathbf{Y}}^{(j+1)}$ with $g'(\mu_{is}^{(j+1)})$ and $V(\mu_{is}^{(j+1)})$, $i=1,\ldots,n$, $s=1,\ldots, t$, using $\bs{\eta}^{(j+1)}$ and $\bs{\mu}^{(j+1)}$.\\
	(iii) Set $j=j+1$.}}
\label{ALGO:PIRLS}
\end{algorithm}
\ULforem}

Compared with the traditional nonparametric techniques, such as kernel smoothing, the proposed algorithm is much more computationally efficient. Therefore, we can easily apply our method to analyze massive spatiotemporal data sets.


\subsection{Modeling the Number of Fatal and Recovered Cases}

To fit the proposed STEM and make predictions for cumulative positive cases, one obstacle is the lack of direct observations for the number of active cases, $I_{it}$. Instead, the most commonly reported number is the count of total confirmed cases, $C_{it}$. Some departments of public health also release information about fatal cases $D_{it}$ and recovered cases $R_{it}$, while such kind of data tends to suffer from missingness, large error and inconsistency due to its difficulty in data collection; see the discussions in \cite{KCRA:20}.

Based on the fact that $I_{it} = C_{it} - R_{it} - D_{it}$, we attempt to modeling $D_{it}$ and $R_{it}$ in order to facilitate the estimation and prediction of newly confirmed cases $Y_{it}$ based on the proposed STEM model. Let $\Delta D_{it} = D_{it}-D_{i,t-1}$ be the new fatal cases on day $t$. Note that patients who die on any given day were infected much earlier. Using similar notations in the STEM model (\ref{model:STEM_infection}), we assume that
\begin{equation}
\Delta D_{it}| \mathbf{X}_i, \mathbf{U}_{i}, I_{i,t-\delta}, \mathbf{A}_{i,t-r}~\sim~\text{Poisson}(\mu^{\D}_{it}),
\label{model:STEM_death}
\end{equation}
where $\delta=14$ is the time delay between illness and death as suggested in \cite{Baud:20}, and
\[
\log(\mu_{it}^{\D})=\beta_{0t}^{\D}(\mathbf{U}_{i})+ \beta_{1t}^{\D}\log(I_{i,t-14})+\sum_{j=1}^p\alpha_{jt}^{\D}A_{ij,t-r}+\sum_{k=1}^q \gamma_{kt}^{\D}(X_{ik}).
\]

Ideally, if sufficient data for recovered cases can be collected from each area, a similar model can be fitted to explain the growth of the recovered cases. However, there are no uniform criteria to collect recovery reports across the US \citep{CNN:20}. According to the CDC, severe cases with COVID-19 often require medical care and receive supportive care in the hospital. At the same time, in general, most people with the mild illness are not hospitalized and suggested to recover at home. Currently, only a few states regularly update the number of recovered patients, but seldom can the counts be mapped to counties. 

Due to the lack of data, we are no longer able to use all the explanatory variables discussed above to model daily new recovered cases. Instead, we mimic the relationship between the number of recovered and active cases from some Compartmental models in epidemiology \citep{Anastassopoulou:etal:20,Siettos:2013}. At current time point $t$, we assume that $\Delta R_{is} = \nu_t I_{i, s-1} + \varepsilon_{is}, ~ s=t-t_0,\ldots,t$, in which the recovery rate $\nu_t$ enables us to make reasonable predictions for future recovered patients counts and provide researchers with the foresight of when the epidemic will end. The rate $\nu_t$ can be either estimated from available state-level data, or obtained from prior medical studies due to the under-reporting issue in actual data.  

\subsection{Zero-inflated Models at the Early Stage of the Outbreak}

It is well known that in the early stage of an epidemic, the quality of any model output can be affected by the restricted quality of data that pertain to under-detection or inconsistent detection of cases, reporting delays, and poor documentation, regarding infections, deaths, tests, and other factors. There are many counties with zero daily counts at the early stage of disease spread. Therefore, we consider zero-inflated models based on a zero-inflated probability distribution, i.e., a distribution that allows for frequent zero-valued observations. Following the previous works \citep{Arab:etal:12,Wood:etal:16}, we assume the observed counts $Y_{it}$ contributes to a zero-inflated Poisson (ZIP) distribution, $\textrm{ZIP}(\mu_{it}^{\I}, p_{it}^{\I})$, specifically, we assume that
\[
P(Y_{it}=y|I_{i,t-1},\mathbf{Z}_{i,t-1}, \mathbf{A}_{i,t-r},\mathbf{X}_i,\mathbf{U}_{i})=
\begin{cases}
1- p_{it}^{\I},& y=0,\\
p_{it}^{\I}\frac{(\mu_{it}^{\I})^y}{\{\exp(\mu_{it}^{\I})-1\}y!},& y>0,
\end{cases}
\]
where $\mu_{it}^{\I}$ follows (\ref{model:STEM_infection}), $p_{it}^{\I}=\mathrm{logit}(\eta_{it}^{\I})$ with $\eta_{it}^{\I} = a_1 + \{b_0+\exp(a_2)\} \log(\mu_{it}^{\I})$ and $a_1, a_2$ are unknown parameters. Here we take $b_0=0$ and $a_1$, $a_2$ are estimated with the roughness parameters. See \cite{Wood:etal:16} for the estimation of $a_1$ and $a_2$.

Let $\Delta D_{it}=D_{it}-D_{i,t-1}$ be the number of new fatal cases on day $t$. Similarly, we can consider zero-inflated models for fatal cases, in which we assume the observed count $\Delta D_{it}$ contributes to a ZIP distribution $\textrm{ZIP}(\mu_{it}^{\D}, p_{it}^{\D})$:
\[
P(\Delta D_{it}=d|I_{i,t-1},\mathbf{A}_{i,t-r}, \mathbf{X}_i,\mathbf{U}_{i})=
\begin{cases}
1- p_{it}^{\D}, & d=0,\\
p_{it}^{\D}\frac{(\mu_{it}^{\D})^d}{\{\exp(\mu_{it}^{\D})-1\}d!}, & d>0,
\end{cases}
\]
where $p_{it}^{\D}=\mathrm{logit}(\eta_{it})$ with $\eta_{it}^{\D}= v_1 + \{c+\exp(v_2)\} \log(\mu_{it}^{\D})$, and $v_1, v_2$ are unknown parameters that can be similarly estimated as in the above.

\setcounter{chapter}{5} 
\setcounter{section}{4} 
\renewcommand{\thesection}{\arabic{section}} %
\renewcommand{\thesubsection}{5.\arabic{subsection}}%
\setcounter{equation}{0} \renewcommand{\theequation}{5.\arabic{equation}} %
\setcounter{table}{0} \renewcommand{\thetable}{{5.\arabic{table}}} %
\setcounter{figure}{0} \renewcommand{\thefigure}{5.\arabic{figure}} %
\setcounter{algorithm}{0} \renewcommand{\thealgorithm}{5.\arabic{algorithm}} %
\setcounter{theorem}{0} \renewcommand{\thetheorem}{{5.\arabic{theorem}}} %
\setcounter{lemma}{0} \renewcommand{\thelemma}{{5.\arabic{lemma}}} %
\setcounter{proposition}{0} \renewcommand{\theproposition}{{5.\arabic{proposition}}}%
\setcounter{corollary}{0} \renewcommand{\thecorollary}{{5.\arabic{corollary}}}%

\section{Forecast and Band of the Forecast Path} \label{sec:prediction}

Understanding the impact of COVID-19 requires accurate forecasting of the spread of infectious cases as well as analysis of the number of deaths and recoveries. In this section, we describe our prediction procedure of these counts; specifically, we are interested in predicting the number of new infected cases and deaths. We also provide the prediction band to quantify the uncertainty of the prediction. 

\subsection{The $h$-step Ahead Prediction}

We first consider an $h$-step ahead prediction. If we observe $C_{is}, I_{is}, R_{is}, D_{is}$ for $s =1,\ldots, t$, then the infection model and fatal cases model can be fitted by regressing $\left\{Y_{is}\right\}_{i=1,s=t-t_0}^{n,t}$, $\left\{\Delta D_{is}\right\}_{i=1,s=t-t_0}^{n,t}$ on $\left\{I_{i,s-1}, Z_{i,s-1}, \mathbf{A}_{i,s-r},\mathbf{X}_i\right\}_{i=1,s =1}^{n,t}$, respectively. In our forecast, we fit the infection model (\ref{model:STEM-infection}) and death model (\ref{model:STEM-death}) alternatively to update the predicted number of cases in each compartment. To be specific, the predictions of at $t+h$ are 
\begin{equation}
\label{EQ:prediction} 
    \widehat{Y}_{i,t+h} = g^{-1}\Big\{\widehat{\beta}_0^{\I} (\mathbf{U}_{i}) + \widehat{\beta}_1^{\I} (\mathbf{U}_{i})\log(\widehat{I}_{i,t+h-1}) + \widehat{\alpha}_0^{\I} \widehat{Z}_{i,t+h-1}
+\sum_{j=1}^p\widehat{\alpha}_j^{\I} A_{ij,t+h-r}+\sum_{k=1}^q \widehat{\gamma}_k^{\I} (X_{ik})\Big\}, 
\end{equation}
where $\widehat{I}_{i,t+h-1} = \widehat{C}_{i,t+h-1} - \widehat{R}_{i,t+h-1} - \widehat{D}_{i,t+h-1}$, $\widehat{C}_{i,t+h-1} = I_{it}+\sum_{j=t+1}^{t+h-1} \widehat{Y}_{ij}$, and $\widehat{Z}_{i,t+h-1} = \log(N_i-\widehat{C}_{i,t+h-1}) - \log(N_i)$.
Meanwhile, let
\[
\widehat{\Delta D}_{i,t+h}  = g^{-1}\Big\{\widehat{\beta}^{\D}_0 (\mathbf{U}_{i}) + \widehat{\beta}^{\D}_1 (\mathbf{U}_{i})\log(I_{i,t+h-\delta})
+\sum_{j=1}^p\widehat{\alpha}^{\D}_jA_{ij,t+h-r^{\prime}}+\sum_{k=1}^q \widehat{\gamma}^{\D}_k(X_{ik})\Big\}, 
\]
and $\widehat{\Delta R}_{i,t+h}  = \widehat{\nu}^{\R} \widehat{I}_{i,t+h-1}$, where we predict $R_{i,t+h}$ by $\widehat{R}_{i, t+h} = R_{it} + \sum_{s=1}^{h}\widehat{\Delta R}_{i,t+s}$, and $D_{i,t+h}$ by $\widehat{D}_{i,t+h} = D_{it} + \sum_{s=1}^h \widehat{\Delta D}_{i,t+s}$. Then, the predicted number of active cases and susceptible cases are $\widehat{I}_{i, t+h} = \widehat{C}_{i,t+h-1} + \widehat{Y}_{i, t+h} - \widehat{R}_{i, t+h} - \widehat{D}_{i, t+h}$, and $\widehat{S}_{i, t+h} = N_{i} -(\widehat{C}_{i,t+h-1} + \widehat{Y}_{i, t+h})$. The above one-step predicted values can be plugged back into equation (\ref{EQ:prediction}) to obtain the predictions for the following days by repeating the same procedure. 

\subsection{Prediction Bands}

There is substantial interest in quantifying the uncertainty for the forecasts with a succession of periods; see, for example, \cite{kong2018prediction} and \cite{IHME:Christopher:20}. To assess the uncertainty associated with the prediction, we propose to develop a prediction band that provides an upper and lower expectation for the real path of the observations.

To construct the band for forecast path $\{Y_{i,t+h}, h=1,\ldots,H\}$, we consider the bootstrap method \citep{Staszewska:09}, in which the bootstrap samples are generated using the bias-corrected bootstrap procedure. Bootstrapping is a powerful and flexible data analysis tool,
which allows us to assign measures of accuracy. By simulating the observed data from our estimated model, we collect the bootstrap samples. For a given bootstrap sample, we re-estimate the model, and produce the corresponding forecast paths. The prediction uncertainty comes from two parts: the estimation variation and the variation from the individual bootstrap sample. We construct $100(1-\alpha)\%$ prediction band based on the envelope covering all the forecast paths after deleting $100\alpha\%$ extreme paths from all bootstrap paths. See Algorithms \ref{ALGO:Bootstrap1} and \ref{ALGO:Bootstrap2} for the details.

\normalem{
\begin{algorithm}
\footnotesize
\SetKwBlock{Begin}{}{}
\caption{\bf{A bootstrap procedure to correct the bias.}}

\BlankLine
\textbf{Step 1.} Fit models 
 (\ref{model:STEM-infection}) and (\ref{model:STEM-death}) using $(Y_{is}, I_{i,s-1}, Z_{i,s-1}, \mathbf{A}_{i,s-r}, \mathbf{X}_{i}, \mathbf{U}_{i})_{i=1,s=1}^{n,t}$ and $(\Delta D_{is}, I_{i,s-1}, \mathbf{A}_{i,s-r}, \mathbf{X}_{i}, \mathbf{U}_{i})_{i=1,s=1}^{n}$, obtain $\widehat{\bs{\beta}}^{\I}$, $\widehat{\bs{\alpha}}^{\I}$, $\widehat{\bs{\gamma}}^{\I}$, $\widehat{\bs{\beta}}^{\D}$, $\widehat{\bs{\alpha}}^{\D}$, $\widehat{\bs{\gamma}}^{\D}$, $\widehat{\nu}^{\R}$, $\widehat{\bs{a}}$, and $\widehat{\bs{v}}$.\\
\Begin(\textbf{Step 2.} Generate bootstrap samples to correct the bias in the estimator of the coefficients.)

{
\ForEach {$1 \leq b \leq B$}{
~(i) Generate the bootstrap sample as follows.

\ForEach {$1 \leq  s \leq t$}{Generate $Y_{is}^b  \sim \textrm{ZIP}(\widehat{\mu}_{is}^{\I}, \widehat{p}^\I_{is})$, $\Delta D_{is}^{b}  \sim \textrm{ZIP}(\widehat{\mu}^{\D}_{is}, \widehat{p}^\D_{is})$, and $\Delta R_{is}^{b} = \widehat{\nu}^{\R} I_{i,s-\delta^{\prime}}$, where
\begin{align*}
	& \widehat{\mu}_{is}^{\I} = \exp\Big\{\widehat{\beta}_{0}^{\I} (\mathbf{U}_{i}) 
		+ \widehat{\beta}^{\I}_{1} (\mathbf{U}_{i})\log(I_{i, s-1})
		+\widehat{\alpha}^{\I}_{0} Z_{i,s-1}
		+\sum_{j=1}^p \widehat{\alpha}^{\I}_{j} A_{ij,s-r}
		+ \sum_{k=1}^q \widehat{\gamma}_{k}^{\I}(X_{ik}) \Big\}, \\[-6pt]
	& \widehat{\mu}_{is}^{\D} = \exp\Big\{\widehat{\beta}_{0}^{\D} (\mathbf{U}_{i}) 
		+ \widehat{\beta}_{1}^{\D} (\mathbf{U}_{i})\log(I_{i, s-\delta})
		+\sum_{j=1}^p \widehat{\alpha}_{j}^{\D} A_{ij,s-r^{\prime}} + \sum_{k=1}^q \widehat{\gamma}_{k}^{\D}(X_{ik}) \Big\}, \\[-4pt]
	&\widehat{p}_{is}^{\I}=\mathrm{logit}\Big[\widehat{a}_1 + \{b_0+\exp(\widehat{a}_2)\} \log(\widehat{\mu}_{is}^{\I})\Big],~ \widehat{p}_{is}^{\D}=\mathrm{logit}\Big[\widehat{v}_1 + \{b_0+\exp(\widehat{v}_2)\} \log(\widehat{\mu}_{is}^{\D})\Big].
\end{align*}
Update $ Z_{is}^b = \log (S_{is}^b/ N_i)$, where $S_{is}^b = S^b_{i, s-1} - Y_{is}^b$ and $I_{is}^b = I_{i,s-1} + Y_{is}^b - \Delta D_{is}^b - \Delta R_{is}^b$. 
} 

\noindent (ii) Fit the models 
 (\ref{model:STEM-infection}) and (\ref{model:STEM-death}) based on $(Y^{b}_{is}, I_{i,s-1}^b, Z_{i,s-1}, \mathbf{A}_{i,s-r}, \mathbf{X}_{i}, \mathbf{U}_{i})_{i=1,s=1}^{n,t}$ and $(\Delta D^{b}_{is}, I_{i,s-\delta}^b, \mathbf{A}_{i,s-r^{\prime}}, \mathbf{X}_{i}, \mathbf{U}_{i})_{i=1,s=1}^{n,t}$, respectively, and obtain $(\widehat{\bs{\beta}}^{\I, b}, \widehat{\bs{\alpha}}^{\I, b}, \widehat{\bs{\gamma}}^{\I, b}, \widehat{\bs{a}}^b)$ and $(\widehat{\bs{\beta}}^{\D, b}, \widehat{\bs{\alpha}}^{\D, b}, \widehat{\bs{\gamma}}^{\D, b},
 \widehat{\bs{v}}^b)$.
}}

\noindent \textbf{Step 3.} Calculate the bias of the coefficients based on the above bootstrap samples. For example, for $\ell=0,1$, let $\text{bias}(\widehat{\beta}^{\I}_{\ell})=B^{-1}\sum_{b=1}^B\widehat{\beta}_{\ell}^{\I, b}-\widehat{\beta}^{\I}_{\ell}$, and let $\widehat{\beta}_{\ell}^{\I, \cc}=\widehat{\beta}^{\I}_{\ell}-\text{bias}(\widehat{\beta}^{\I}_{\ell})$ be the corrected coefficient function. Similarly, we obtain the bias-corrected coefficients of $\widehat{\bs{\alpha}}_{t}$,  $\widehat{\bs{\gamma}}_{t}$, $\widehat{\bs{a}}$, and $\widehat{\bs{v}}$, denoted by $\widehat{\bs{\alpha}}_{t}^{\cc}$, $\widehat{\bs{\gamma}}_{t}^{\cc}$, $\widehat{\bs{a}}^{\cc}$, and $\widehat{\bs{v}}^{\cc}$, respectively. 
\vskip .2in
\label{ALGO:Bootstrap1}
\end{algorithm}
\ULforem}

\normalem{
\begin{algorithm}
\footnotesize
\SetKwBlock{Begin}{}{}

\caption{{\bf A bootstrap procedure to calculate the prediction band.}}
\BlankLine
\Begin(\textbf{Step 1.} Generate bootstrap samples to construct prediction band.)

{
\ForEach {$1 \leq b \leq B$}{
\ForEach {$1 \leq  h \leq H$}{Generate $Y_{i,t+h}^b  \sim \textrm{ZIP}(\widehat{\mu}_{i,t+h}^{\I,\cc,b}, \widehat{p}_{i,t+h}^{\I,\cc,b})$, $\Delta D_{i,t+h}^{b}  \sim \textrm{ZIP}(\widehat{\mu}^{\D,\cc, b}_{i,t+h}, \widehat{p}_{i,t+h}^{\D,\cc,b})$, and $\Delta R_{i, t+h}^{b} = \widehat{\nu}^{\R} I_{i,t+h-\delta^{\prime}}$ based on bootstrap estimators, where
\begin{align*}
& \widehat{\bs{\beta}}^{\I, \cc, b} = 2 \widehat{\bs{\beta}}^{\I} - \widehat{\bs{\beta}}^{\I, b},~ \widehat{\bs{\alpha}}^{\I, \cc, b} = 2 \widehat{\bs{\alpha}}^{\I} - \widehat{\bs{\alpha}}^{\I, b},~\widehat{\bs{\gamma}}^{\I, \cc, b} = 2 \widehat{\bs{\gamma}}^{\I} - \widehat{\bs{\gamma}}^{\I, b}, ~\widehat{\bs{a}}^{\cc, b} = 2 \widehat{\bs{a}} - \widehat{\bs{a}}^{b}\\[-3pt]
&\widehat{\bs{\beta}}^{\D, \cc, b} = 2 \widehat{\bs{\beta}}^{\D} - \widehat{\bs{\beta}}^{\D, b},~ \widehat{\bs{\alpha}}^{\D, \cc, b} = 2 \widehat{\bs{\alpha}}^{\D} - \widehat{\bs{\alpha}}^{\D, b},~\widehat{\bs{\gamma}}^{\D, \cc, b} = 2 \widehat{\bs{\gamma}}^{\D} - \widehat{\bs{\gamma}}^{\D, b},~\widehat{\bs{v}}^{\cc, b} = 2 \widehat{\bs{v}} - \widehat{\bs{v}}^{b},\\[-3pt]
& \widehat{\mu}_{i,t+h}^{\I, \cc, b} = \exp\Big\{\widehat{\beta}_0^{\I, \cc, b} (\mathbf{U}_{i}) 
		+ \widehat{\beta}_1^{\I, \cc, b} (\mathbf{U}_{i})\log(I_{i, t+h-1})
		+\widehat{\alpha}_0^{\I, \cc, b} Z_{i,t+h-1}\\[-8pt]
		&\qquad \qquad +\sum_{j=1}^p \widehat{\alpha}_j^{\I, \cc, b} A_{ij,t+h-r} + \sum_{k=1}^q \widehat{\gamma}_k^{\I, \cc, b}(X_{ik}) \Big\}, \\[-8pt]
& \widehat{\mu}_{i,t+h}^{\D,\cc, b} = \exp\Big\{\widehat{\beta}_0^{\D,\cc,b} (\mathbf{U}_{i}) 
		+ \widehat{\beta}_1^{\D,\cc,b} \log(I_{i, t+h-\delta})
		+\sum_{j=1}^p \widehat{\alpha}_j^{\D,\cc, b} A_{ij,t+h-r^{\prime}} + \sum_{k=1}^q \widehat{\gamma}_k^{\D,\cc, b}(X_{ik}) \Big\},\\[-6pt]
&\widehat{p}_{i,t+h}^{\I,\cc,b}=\mathrm{logit}\Big[\widehat{a}_1^{\cc,b} + \{b_0+\exp(\widehat{a}_2^{\cc,b})\} \log(\widehat{\mu}_{i,t+h}^{\I, \cc, b})\Big],\\[-3pt]
&\widehat{p}_{i,t+h}^{\D,\cc,b}=\mathrm{logit}\Big[\widehat{v}_1^{\cc,b} + \{b_0+\exp(\widehat{v}_2^{\cc,b})\} \log(\widehat{\mu}_{i,t+h}^{\D, \cc, b})\Big].
\end{align*}
Update $Z_{i,t+h}^b = \log (S_{i,t+h}^b/ N_i)$, where $S_{i,t+h}^b = S^b_{i, t+h - 1} - Y_{i,t+h}^b$ and $I_{i,t+h}^b = I_{i,t+h-1}^b + Y_{i,t+h}^b - \Delta D_{i,t+h}^b - \Delta R_{i,t+h}^b$. 
}}}

\textbf{Step 2.}  Construct the $100(1-\alpha)\%$ prediction band by the above $B$ bootstrap paths with the most extreme $\alpha B$ paths discarded. Start with setting $\kappa=0$.

\While{$\kappa<\alpha B$}{

(i) For each forecast time point $h=1,\ldots, H$ (there are in total $B-\kappa$ constructed paths available), identify the largest and the smallest bootstrap forecast values, and the associated paths. Notice there are $2H$ extreme values and at most corresponding $2H$ paths.

(ii) Compute the distances from each of the bootstrap path (at most $2H$) to the bootstrap sample, based on: $\sum_{h=1}^{H}(\widehat{\mu}_{i,t+h}^{\cc}-Y_{i,t+h}^{b})^{2}$ or $\sum_{h=1}^{H}|\widehat{\mu}_{i,t+h}^{\cc}-Y_{i,t+h}^{b}|$. 

(iii) Remove the path with the largest distance, and set $\kappa=\kappa+1$.
}

\textbf{Step 3.} Obtain the $100(1-\alpha)\%$ prediction band from the envelope of the remaining $(1-\alpha)B$ paths.

\vskip .2in
\label{ALGO:Bootstrap2}
\end{algorithm}
\ULforem}

\setcounter{chapter}{6} 
\setcounter{section}{5} 
\renewcommand{\thesection}{\arabic{section}} %
\renewcommand{\thesubsection}{6.\arabic{subsection}}%
\setcounter{equation}{0} \renewcommand{\theequation}{6.\arabic{equation}} %
\setcounter{table}{0} \renewcommand{\thetable}{{6.\arabic{table}}} %
\setcounter{figure}{0} \renewcommand{\thefigure}{6.\arabic{figure}} %
\setcounter{algorithm}{0} \renewcommand{\thealgorithm}{6.\arabic{algorithm}} %
\setcounter{theorem}{0} \renewcommand{\thetheorem}{{6.\arabic{theorem}}} %
\setcounter{lemma}{0} \renewcommand{\thelemma}{{6.\arabic{lemma}}} %
\setcounter{proposition}{0} \renewcommand{\theproposition}{{6.\arabic{proposition}}}%
\setcounter{corollary}{0} \renewcommand{\thecorollary}{{6.\arabic{corollary}}}%

\section{A Simulation Study}
\label{sec:simulation}

\begin{figure}[!htbp]
\begin{center}
\begin{tabular}{ccc}
\includegraphics[height=1.5in,width=1.75in]{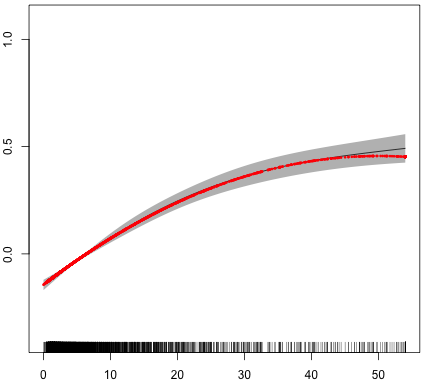} &
\includegraphics[height=1.5in,width=1.75in]{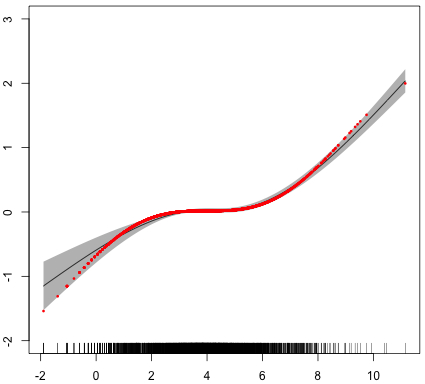} &
\includegraphics[height=1.5in,width=1.75in]{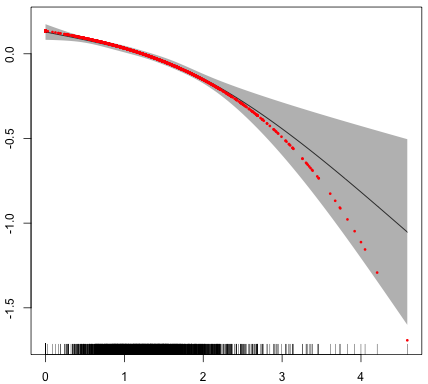} \\
(a) $\gamma_1^{\I}(\cdot)$ & (b) $\gamma_2^{\I}(\cdot)$ & (c) $\gamma_3^{\I}(\cdot)$ \\
\includegraphics[height=1.5in,width=1.75in]{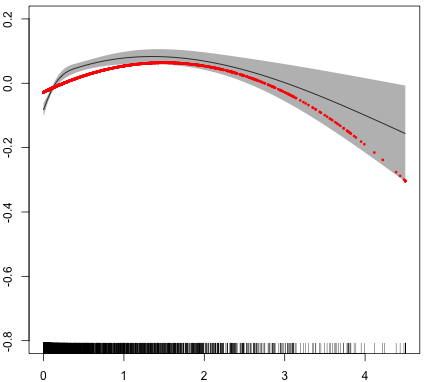} & 
\includegraphics[height=1.5in,width=1.75in]{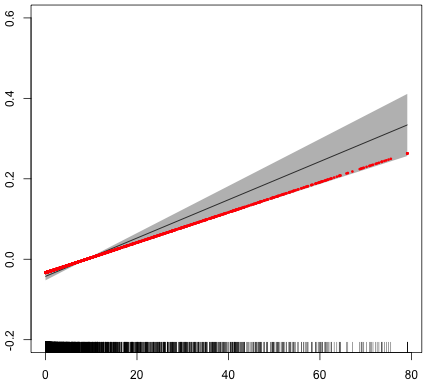} \\
(d) $\gamma_4^{\I}(\cdot)$ & (e) $\gamma_5^{\I}(\cdot)$\\
\end{tabular}
\end{center}
\caption{The true and estimated univariate component functions in model (\ref{model:STEM-infect-sim}) (estimation window: 04/17/20-04/30/20). The red curves represent true functions, while black curves and dark area indicate estimated coefficients and their 95\% confidence bands.}
\label{FIG:sim_uni}
\end{figure}

\begin{figure}[htbp]
\begin{center}
\renewcommand{\arraystretch}{0.75}
\begin{tabular}{ccc}
\includegraphics[width = 0.3\textwidth]{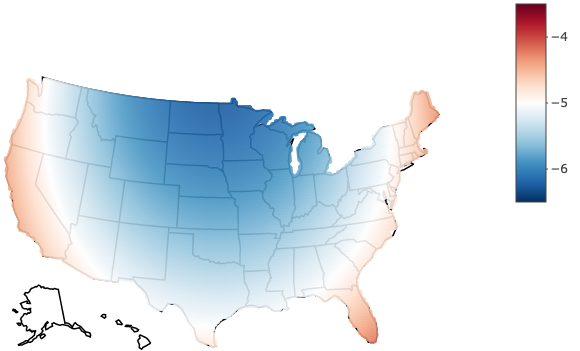} &
\includegraphics[width = 0.3\textwidth]{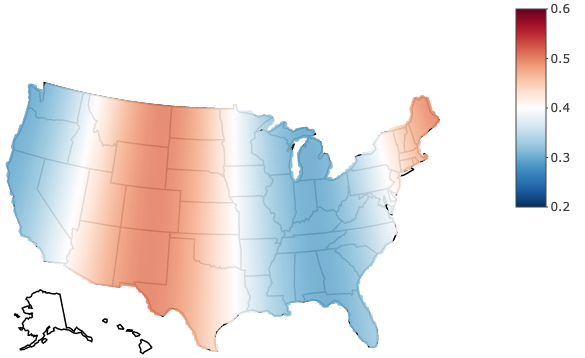} &
\includegraphics[width = 0.3\textwidth]{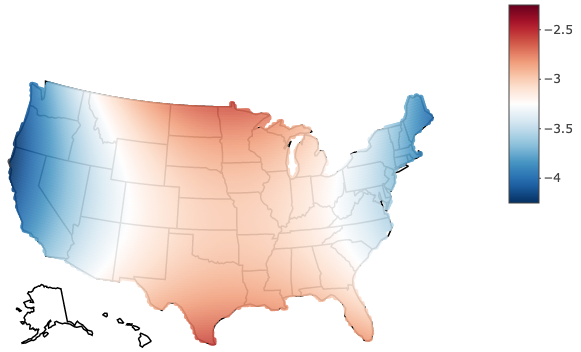}\\
(a) $\beta_{0}^{\I}(\cdot)$  & (b) $\beta_{1}^{\I}(\cdot)$ & (c) $\beta_{0}^{\D}(\cdot)$\\
\includegraphics[width = 0.3\textwidth]{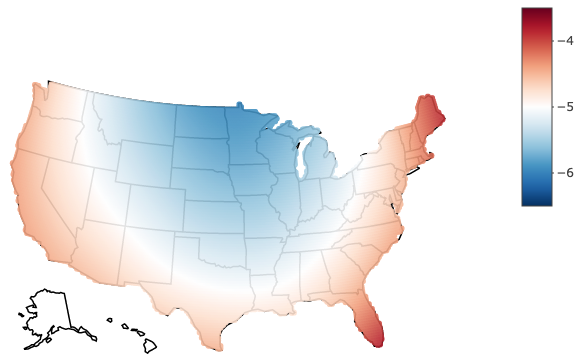} 
& \includegraphics[width = 0.3\textwidth]{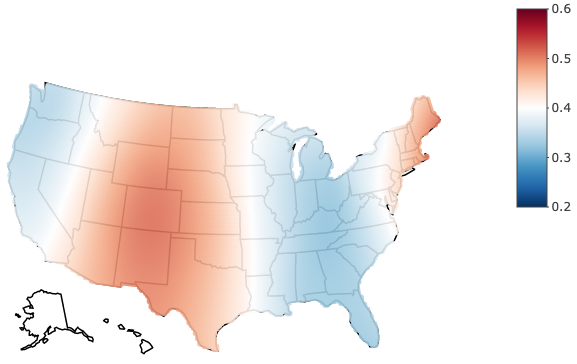}
& \includegraphics[width = 0.3\textwidth]{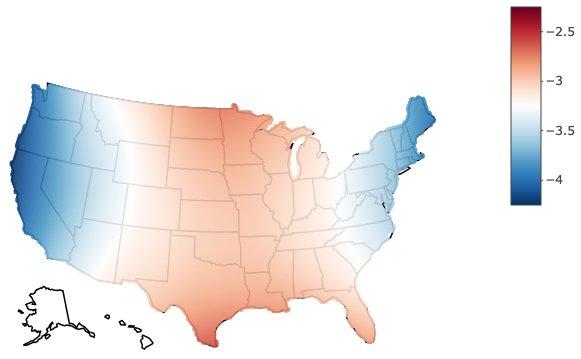}\\
(d) $\widehat{\beta}_{0}^{\I}(\cdot)$  & (e) $\widehat{\beta}_{1}^{\I}(\cdot)$ & (f) $\widehat{\beta}_{0}^{\D}(\cdot)$\\
\end{tabular}
\end{center}
\caption{True and estimated bivariate varying coefficients (estimation window: 04/17/20-04/30/20) in the simulation.}
\label{FIG:sim_beta}
\end{figure}

In this section, we conduct a simulation study to evaluate the finite sample performance of the proposed method. In the simulation, we use a subset of covariates of the county-level characteristics analyzed in Section \ref{sec:application}. The response variable $Y_{it}$ and $\Delta D_{it}$ are generated from a ZIP distribution with the logarithm of Poisson parameters generated as following: 
\begin{align}
    \log(\mu_{it}^{\I}) &= \beta_{0}^{\I} (\mathbf{U}_i) + \beta_{1}^{\I} (\mathbf{U}_i) \log(I_{i,t-1}) +\sum_{j=1}^p\alpha_{j}^{\I} A_{ij,t-r}+\sum_{k=1}^q \gamma_{k}^{\I}(X_{ik}), \label{model:STEM-infect-sim}\\
    \log(\mu_{it}^{\D}) &=\beta_{0}^{\D} (\mathbf{U}_i) + \beta_{1}^{\D}\log(I_{i,t-\delta}), \label{model:STEM-death-sim}
\end{align}
where $\delta = 14$, $p = 2$, $q = 5$, $r = 7$, and $A_{ijt}$, $X_{ik}$, $j = 1,2$, $k=1,\ldots,5$, come from the covariates in the COVID-19 dataset described in Section \ref{sec:data}. The true univariate functions $\gamma_1(x)$, $\gamma_2(x)$, \ldots, $\gamma_{5}(x)$, together with their estimate and confidence bands in one typical iteration, are displayed in Figure~\ref{FIG:sim_uni} (a)--(e). Figure \ref{FIG:sim_beta} (a)--(c) depict the bivariate coefficient functions $\beta_{0}^{\I}(\cdot)$, $\beta_{1}^{\I}(\cdot)$ and $\beta_{0}^{\D}(\cdot)$, which are generated to mimic the spatial pattern of infection/mortality rate in the pandemic. We also include the corresponding estimated functions from one experiment in Figure \ref{FIG:sim_beta} (d)--(f) to show that the spatial pattern can be very well captured using the proposed method. For recovery data, the daily recovered cases are simulated by $\Delta R_{it} = \nu^{\R} I_{i, s-1}$, where $\nu^{\R} = 0.07$. We simulate data by assuming that a pandemic emerged on March 15 with $1$ case showed up in each of the $420$ selected counties. These counties are selected if COVID-19 cases had been found by March 15 in real data. Then, daily confirmed, fatal and recovered cases are generated based on model (\ref{model:STEM-infect-sim}) and (\ref{model:STEM-death-sim}) from a ZIP distribution with the complimentary log of the zero probability being linearly dependent on the log of the Poisson parameter $\mu_{it}^{\I}$ and $\mu_{it}^{\D}$. 

\begin{figure}[htbp]
\begin{center}
\renewcommand{\arraystretch}{0.75}
\begin{tabular}{cc}
\includegraphics[height=1.25in]{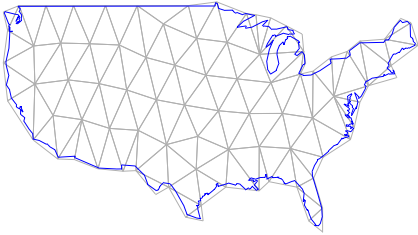} & \includegraphics[height=1.25in]{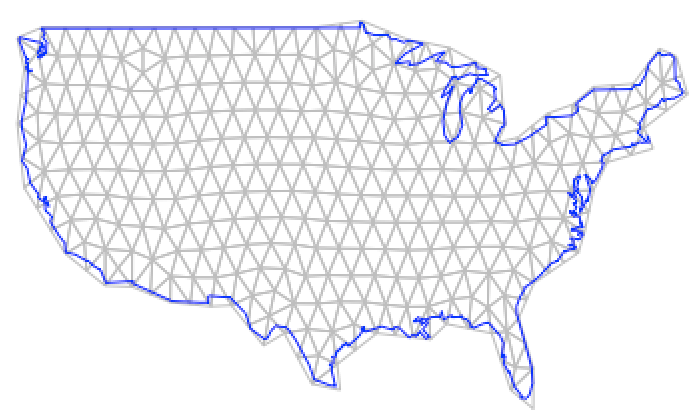}\\
(a) triangulation $\triangle_1$  & (b)  triangulation $\triangle_2$\\
\end{tabular}
\end{center} 
\caption{Triangulations used in the bivariate spline estimation.} 
\label{FIG:triangulations}
\end{figure}

To evaluate the performance numerically, we conduct $100$ Monte Carlo experiments with $9$ or $14$ days as the training window sizes. For the univariate spline smoothing, we use cubic splines with two interior knots; and for the bivariate spline smoothing, we consider degree $d=2$, smoothness $r=1$, and two different triangulations in Figure \ref{FIG:triangulations}: $\triangle_1$ (119 triangles with 87 vertices) and $\triangle_2$ (522 triangles with 306 vertices). The root mean squared error (RMSE) for the parametric and nonparametric components in models (\ref{model:STEM-infect-sim}) and (\ref{model:STEM-death-sim}) are reported in Table \ref{TAB:RMSE_sim}. According to the numeric results, the proposed model is not sensitive to the choice of triangulation. However, increasing the window size of training data can help improve the accuracy in estimating most of the coefficient functions. 

\begin{table}
\caption{\label{TAB:RMSE_sim}
The average of root mean squared errors (RMSEs) of the estimated components in infection model and death model in the simulation (``--'' indicates not applicable).}
\centering
\scalebox{0.85}{
\begin{tabular}{ccccccccccc} \hline\hline
\multirow{2}{*}{Model} & Window &\multicolumn{9}{c}{Component} \\ \cline{3-11}
& Size &$\beta_0$ & $\beta_1$ & $\alpha_1$ & $\alpha_2$ & $\gamma_1$ &$\gamma_2$ & $\gamma_3$ & $\gamma_4$ & $\gamma_5$\\\hline
\multirow{2}{*}{Infection} & 9 days & 0.5094 & 0.0453 & 0.0534 & 0.0258 & 0.0105 & 0.0415 & 0.0214 & 0.0262 & 0.0114\\
& 14 days & 0.4862 & 0.0395 & 0.0499 & 0.0193 & 0.0103 & 0.0416 & 0.0186 & 0.0256 & 0.0119\\
\multirow{2}{*}{Death} & 9 days & 0.0456 & 0.0999 & -- & -- & --& -- & -- & -- & -- \\
& 14 days & 0.0373 & 0.1000 & -- & -- & --& -- & -- & -- & -- \\\hline\hline
\end{tabular}}
\end{table}

In addition, the root mean squared prediction error for $h$-day ahead predictions using $14$ days as the training period are presented in Table \ref{TAB:RMSPE_sim}, $h = 1, \ldots, 28$. For comparison, we also consider the naive model that assumes a linear growth pattern for total confirmed cases for each county that $\mathrm{E}(C_{it}|t) = \beta_{i0} + \beta_{i1}t$,  $\mathrm{Var}(C_{it}|t) = \sigma_{i}^{2}$, where $C_{it}$ stands for the number of total confirmed cases in $i$th county. Comparing with the linear model separately fitted for each county without taking into account the spatial correlation, the prediction performance of STEM is better, especially for the long-term forecast. We further construct $95\%$ prediction bands based on the proposed bootstrap method and calculate the empirical coverage rate based on $100$ replications. Table \ref{Simu1_coverage} reports summary statistics of the coverage rates for $3104$ counties in $48$ contiguous states. For each county, we examine how often, out of $100$ iterations, that the true infected/fatal cases curve falls inside the band all the time for a two-week or four-week forecasting period. Eventually, the coverage rates are very close to the nominal level for most counties, as shown by the quartiles.

\renewcommand{\baselinestretch}{1.25}
\begin{table}[h!tb]
\begin{center}
\caption{The average of root mean squared prediction errors ($\text{RMSPE}$) of the infection or death count, where D$_h$ is for the $h$-day ahead prediction, $h=1,\ldots,28$.}
\label{TAB:RMSPE_sim}
\scalebox{0.8}{
\begin{tabular}{lrrrrrrrrrr}
	\hline \hline
\multicolumn{11}{c}{Infection Model}\\ \hline
Model &$\text{D}_1$ &$\text{D}_2$ &$\text{D}_3$ &$\text{D}_4$ &$\text{D}_5$ &$\text{D}_6$ &$\text{D}_7$ &$\text{D}_8$ &$\text{D}_9$ &$\text{D}_{10}$ \\ \hline
STEM & 3.28 & 5.01 & 6.76 & 8.36 & 9.93 & 11.55 & 13.58 & 15.32 & 17.15 & 18.94  \\ 
Linear & 7.11 & 9.53 & 11.99 & 14.57 & 17.76 & 21.10 & 24.88 & 29.24 & 34.08 & 39.58 \\ \hline
Model & $\text{D}_{11}$ &$\text{D}_{12}$ &$\text{D}_{13}$ &$\text{D}_{14}$ & $\text{D}_{15}$ &$\text{D}_{16}$ &$\text{D}_{17}$ &$\text{D}_{18}$ &$\text{D}_{19}$ &$\text{D}_{20}$ \\ \hline
STEM & 20.91 & 22.83 & 24.78 & 26.68 & 28.76 & 30.86 & 32.93 & 35.20 & 37.34 & 39.54 \\ 
Linear & 45.71 & 52.23 & 58.98 & 66.09 & 73.27 & 80.55 & 88.03 & 95.72 & 103.75 & 112.25 \\ \hline
Model& $\text{D}_{21}$ &$\text{D}_{22}$ &$\text{D}_{23}$ &$\text{D}_{24}$ &$\text{D}_{25}$ &$\text{D}_{26}$ &$\text{D}_{27}$ &$\text{D}_{28}$\\ \hline
STEM & 41.77 & 43.94 & 46.17 & 48.47 & 52.68 & 55.42 & 58.27 & 61.22 \\
Linear & 121.27 & 130.83 & 140.92 & 151.77 & 166.06 & 181.3 & 197.78 & 215.55 \\
\hline
\multicolumn{11}{c}{Death Model}\\ \hline
Model &$\text{D}_1$ &$\text{D}_2$ &$\text{D}_3$ &$\text{D}_4$ &$\text{D}_5$ &$\text{D}_6$ &$\text{D}_7$ &$\text{D}_8$ &$\text{D}_9$ &$\text{D}_{10}$ \\ \hline
STEM & 1.17 & 1.67 & 2.05 & 2.40 & 2.7 & 2.96 & 3.21 & 3.45 & 3.68 & 3.90 \\
Linear & 2.43 & 3.16 & 3.91 & 4.67 & 5.47 & 6.31 & 7.20 & 8.13 & 9.10 & 10.08 \\ \hline
Model &$\text{D}_{11}$ &$\text{D}_{12}$ &$\text{D}_{13}$ &$\text{D}_{14}$ &$\text{D}_{15}$ &$\text{D}_{16}$ &$\text{D}_{17}$ &$\text{D}_{18}$ &$\text{D}_{19}$ &$\text{D}_{20}$ \\
\hline
STEM & 4.08 & 4.29 & 4.49 & 4.69 & 4.87 & 5.04 & 5.18 & 5.33 & 5.47 & 5.60 \\
Linear & 11.06 & 12.11 & 13.21 & 14.32 & 15.46 & 16.66 & 17.84 & 19.11 & 20.39 & 21.69 \\ \hline
Model &$\text{D}_{21}$ &$\text{D}_{22}$ &$\text{D}_{23}$ &$\text{D}_{24}$ &$\text{D}_{25}$ &$\text{D}_{26}$ &$\text{D}_{27}$ &$\text{D}_{28}$\\ \hline
STEM & 5.78 & 5.93 & 6.10 & 6.26 & 6.42 & 6.59 & 6.76 & 6.95\\
Linear & 23.02 & 24.38 & 25.81 & 27.26 & 28.78 & 30.3 & 31.89 & 33.53\\
\hline \hline
\end{tabular}}
\end{center} 
\end{table}

\begin{table}
\caption{\label{Simu1_coverage}
The empirical coverage rate of the 95\% projection band.}
\centering
\scalebox{0.8}{
\begin{tabular}{ccccccc}
\multicolumn{3}{c}{Infection Model} & & \multicolumn{3}{c}{Death Model}\\\cline{1-3} \cline{5-7}
\multirow{2}{*}{Quantile} & \multicolumn{2}{c}{Number of prediction days} && \multirow{2}{*}{Quantile} & \multicolumn{2}{c}{Number of prediction days}\\ \cline{2-3} \cline{6-7}
& 14 days & 28 days &&& 14 days & 28 days\\
\cline{1-3} \cline{5-7}
Q1 & 95\% & 90\% && Q1 & 96\% & 93\% \\
Q2 & 97\% & 93\% && Q2 & 97\% & 95\% \\
Q3 & 98\% & 95\% && Q3 & 98\% & 96\% \\
\cline{1-3} \cline{5-7}
		\end{tabular}}
\end{table}

\setcounter{chapter}{7} 
\setcounter{section}{6} 
\renewcommand{\thesection}{\arabic{section}} %
\renewcommand{\thesubsection}{7.\arabic{subsection}}%
\setcounter{equation}{0} \renewcommand{\theequation}{7.\arabic{equation}} %
\setcounter{table}{0} \renewcommand{\thetable}{{7.\arabic{table}}} %
\setcounter{figure}{0} \renewcommand{\thefigure}{7.\arabic{figure}} %
\setcounter{algorithm}{0} \renewcommand{\thealgorithm}{7.\arabic{algorithm}}%
\setcounter{theorem}{0} \renewcommand{\thetheorem}{{7.\arabic{theorem}}} %
\setcounter{lemma}{0} \renewcommand{\thelemma}{{7.\arabic{lemma}}} %
\setcounter{proposition}{0} \renewcommand{\theproposition}{{7.\arabic{proposition}}}%
\setcounter{corollary}{0} \renewcommand{\thecorollary}{{7.\arabic{corollary}}}%

\section{Analysis and Findings in COVID-19} \label{sec:application}
The goals of the following study are two-fold. First, we develop a new dynamic epidemic modeling framework for public health surveillance data to study the spatiotemporal pattern in the spread of COVID-19. We aim to investigate whether the proposed model could guide the modeling of the dynamics of the spread at the county level by moving beyond the typical theoretical conceptualization of context where a county's infection is only associated with its own features. Second, to understand the factors that contribute to the spread of COVID-19, we model the daily infected cases at the county level, considering the demographic, environmental, behavioral, and socioeconomic factors in the US.

\subsection{Estimation and Inference for the STEM} 

For the model estimation, we consider the following model for the infected count:
\begin{align}
	\log(\mu_{it}^{\I}) 
	=& \beta_{0t}^{\I}(\mathbf{U}_{i}) + \beta_{1t}^{\I}(\mathbf{U}_{i})\log(I_{i,t-1}) +\alpha_{0t}^{\I} Z_{i, t-1} + \alpha_{1t}^{\I}  \mathrm{Control}_{i,t-7} + \alpha_{2t}^{\I}  \mathrm{Mobility}_{i,t-7} \nonumber\\
	& + \gamma_{1t}^{\I}(\mathrm{Gini}_i) 
	+ \gamma_{2t}^{\I}(\mathrm{Affluence}_i) + \gamma_{3t}^{\I}(\mathrm{Disadvantage}_i) +
	\gamma_{4t}^{\I}(\mathrm{Urban}_i) + \gamma_{5t}^{\I}(\mathrm{PD}_i)\nonumber\\
	& + \gamma_{6t}^{\I}(\mathrm{Tbed}_i)
	 + \gamma_{7t}^{\I}(\mathrm{NHIC}_i)     
	 + \gamma_{8t}^{\I}(\mathrm{EHPC}_i) \nonumber\\
	& + \gamma_{9t}^{\I}(\mathrm{AA}_i) 
	+ \gamma_{10t}^{\I}(\mathrm{HL}_i)  
	+ \gamma_{11t}^{\I}(\mathrm{Sex}_i) 
	+ \gamma_{12t}^{\I}(\mathrm{Old}_i), \label{model:infection}
\end{align}
where $i=1,\ldots, 3104$. For the death count, we consider the following semiparametric model: 
\begin{align}
	\log&(\mu_{it}^{\D}) = \beta_{0t}^{\D}(\mathbf{U}_{i}) 
	+ \beta_{1t}^{\D}\log(I_{i,t-\delta})
    + \alpha_{1t}^{\D} \mathrm{Control}_{i,t-7} 
    + \alpha_{2t}^{\D}  \mathrm{Mobility}_{i,t-7} \nonumber\\
	& + \gamma_{1t}^{\D}\mathrm{Gini}_i 
	+\gamma_{2t}^{\D}\mathrm{Affluence}_i 
	+ \gamma_{3t}^{\D}\mathrm{Disadvantage}_i
	+ \gamma_{4t}^{\D}\mathrm{Urban}_i
	+ \gamma_{5t}^{\D}\mathrm{PD}_i   
	  \nonumber\\
	& + \gamma_{6t}^{\D}\mathrm{Tbed}_i +\gamma_{7t}^{\D}\mathrm{NHIC}_i 
	 + \gamma_{8t}^{\D}\mathrm{EHPC}_i
	 + \gamma_{9t}^{\D}\mathrm{AA}_i 
     + \gamma_{10t}^{\D}\mathrm{HL}_i 
     + \gamma_{11t}^{\D}\mathrm{Sex}_i 
     + \gamma_{12t}^{\D}\mathrm{Old}_i.
	\label{model:death}
\end{align}

We consider the data collected from March 16 to September 3, 2020; see the data description in Section \ref{sec:data}. Note that in models (\ref{model:infection}) and (\ref{model:death}), the covariate $\mathrm{Control}_{it}$ is a dummy variable for the executive order ``shelter-in-place'' or ``stay-at-home'', namely $\mathrm{Control}_{it}=1$ suggesting ``shelter-in-place'' taken place for county $i$ at time $t$, while $\mathrm{Control}_{it}=0$ suggesting no restriction or restriction lifted. See Table \ref{Tab:covariates} for details of other county-level predictors.

We use $28$ days, two incubation periods, as an estimation window to examine how the covariates affect the newly infected cases and fatal cases, and we choose $\delta=14$. The roughness parameters are selected by the generalized cross-validation (GCV). The performance of the univariate and bivariate splines is dependent upon the choice of the knots and triangulations, respectively. We use cubic splines with two interior knots for the univariate spline smoothing. We generate the triangulations according to ``max-min" criterion, which maximizes the minimum angle of all the angles of the triangles in the triangulation. We consider the same  triangulations as shown in Figure \ref{FIG:triangulations}: $\triangle_1$ and $\triangle_2$. By the ``max-min" criterion, $\triangle_2$ is better than $\triangle_1$, but it also significantly increases the number of parameters to estimate. As a trade-off, for the estimation of $\beta_0^{\I}(\cdot)$ and $\beta_1^{\I}(\cdot)$, we adopt the finer triangulation $\triangle_2$, and use the rough triangulation $\triangle_1$ to estimate $\beta_0^{\D}(\cdot)$ due to the sparsity problem in the death count and many zeros observed.

\begin{figure}[htbp]
\begin{center}
\renewcommand{\arraystretch}{0.3}
\begin{tabular}{cc}
	\includegraphics[height=1.6in,width=2.4in]{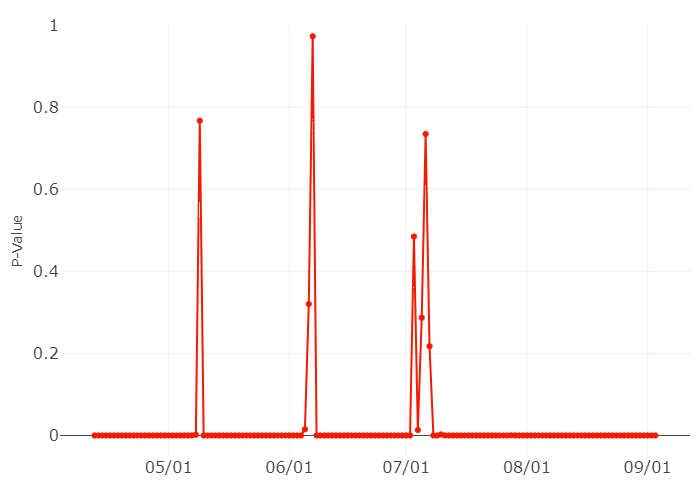} &\includegraphics[height=1.6in,width=2.4in]{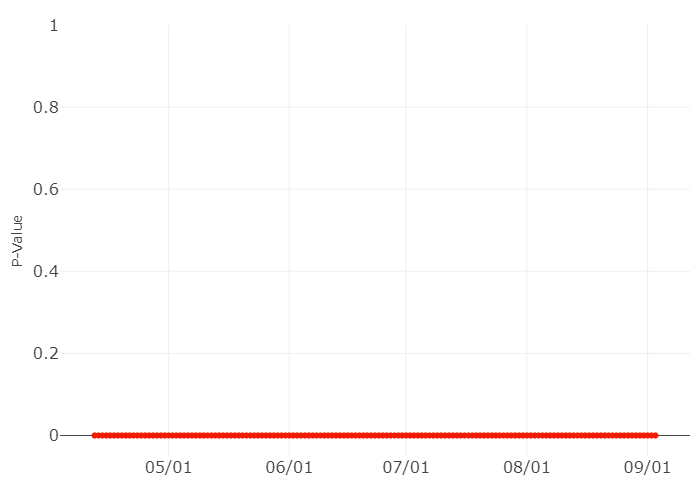}\\[-3pt]
	(a) Control ($\alpha_{1t}^{\I}$) &(b) Mobility ($\alpha_{2t}^{\I}$)
\end{tabular}
\end{center}
\caption{P-values of hypothesis tests of constant coefficients in model (\ref{model:infection}).}
\label{FIG:Infect-p-values}
\end{figure}

First of all, we describe our findings from modeling the COVID-19 related infection counts in 3,104 counties from the 48 mainland US states and the District of Columbia. To examine the effect of the control measures (``shelter-in-place'' or ``stay-at-home'' orders) and mobility level after 7 days, we test the hypothesis: $H_0: \alpha_{jt}^{\I}=0,~j=1,2$ in model (\ref{model:infection}). Figure \ref{FIG:Infect-p-values} (a) shows that the control measure is significant for the infected count most of the time. Figure \ref{FIG:Infect-p-values} (b) shows that the $p$-value of the mobility is always less than 0.0001, and thus the mobility is significant for the entire study period.

\begin{figure}[htbp]
\begin{center}
\scalebox{0.85}{
\begin{tabular}{cccc}
\includegraphics[width = 0.28\textwidth]{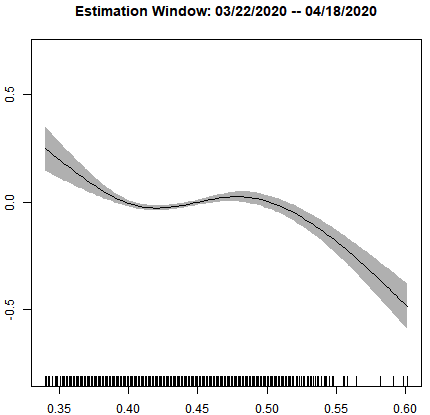} &
\includegraphics[width = 0.28\textwidth]{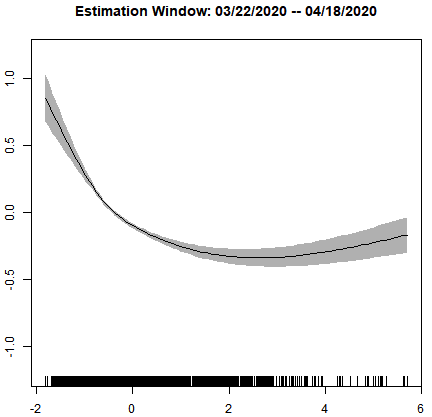} &
\includegraphics[width = 0.28\textwidth]{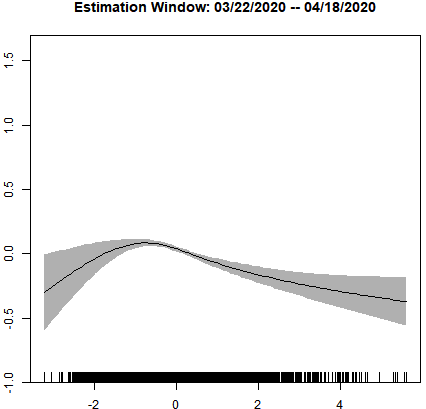} & \includegraphics[width = 0.28\textwidth]{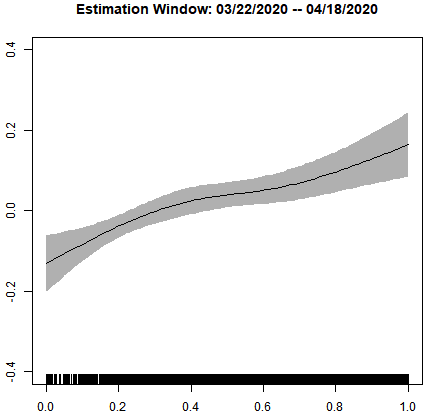}\\
(a) Gini ($\gamma_{1t}^{\I}$) & (b) Affluence ($\gamma_{2t}^{\I}$) & (c) Disadvantage ($\gamma_{3t}^{\I}$) & (d) Urban ($\gamma_{4t}^{\I}$)\\
\includegraphics[width = 0.28\textwidth]{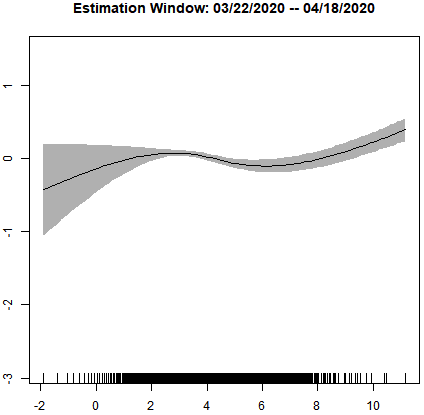} &
\includegraphics[width = 0.28\textwidth]{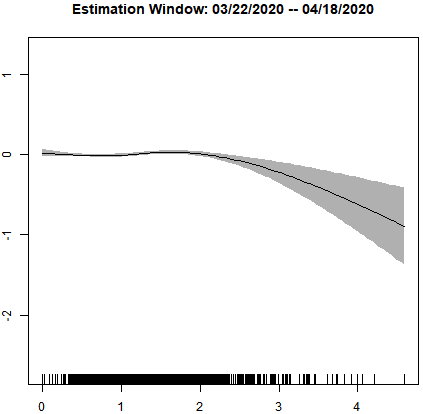} & \includegraphics[width = 0.28\textwidth]{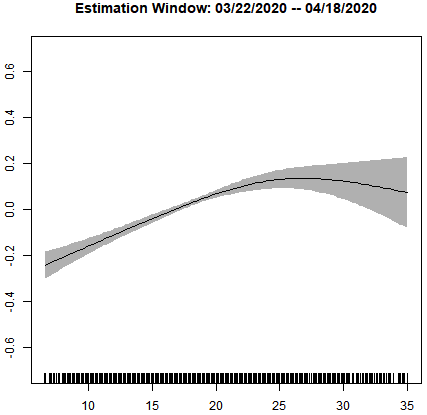} &
\includegraphics[width = 0.28\textwidth]{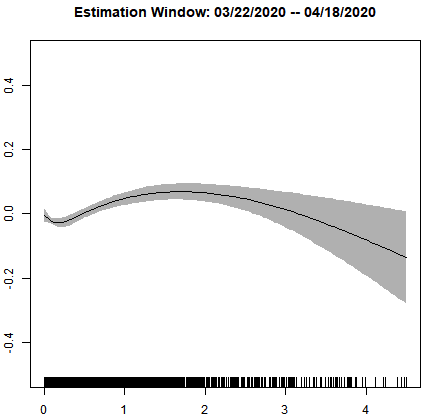}\\
(e) PD ($\gamma_{5t}^{\I}$) & (f) Tbed ($\gamma_{6t}^{\I}$) & (g) NHIC ($\gamma_{7t}^{\I}$) & (h) EHPC ($\gamma_{8t}^{\I}$)\\
\includegraphics[width = 0.28\textwidth]{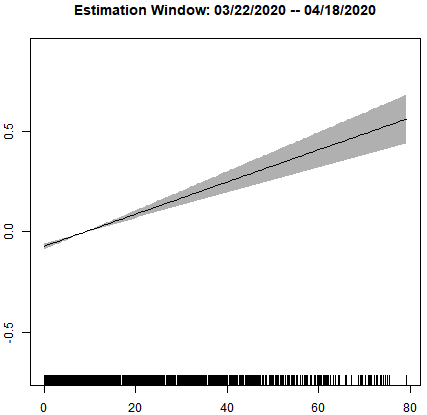} & 
\includegraphics[width = 0.28\textwidth]{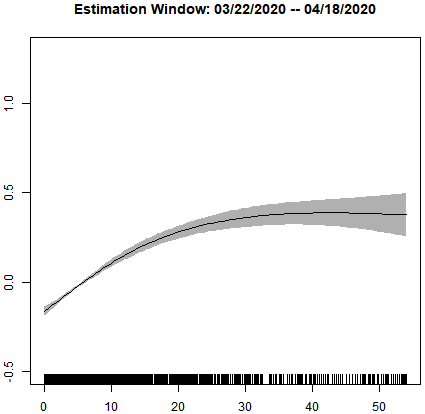} &
\includegraphics[width = 0.28\textwidth]{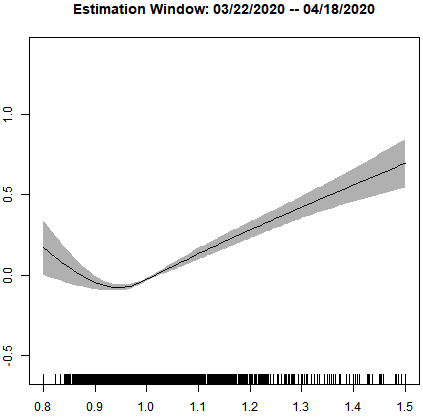} &
\includegraphics[width = 0.28\textwidth]{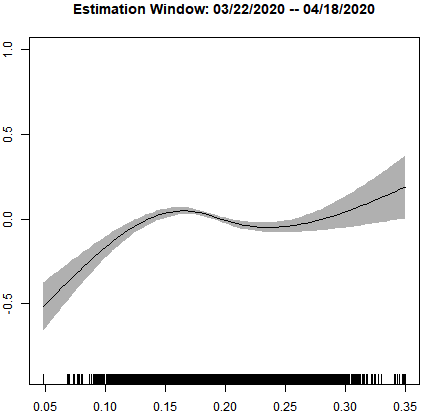}\\
(i) AA ($\gamma_{9t}^{\I}$) & (j) HL ($\gamma_{10t}^{\I}$) & (k) SEX ($\gamma_{11t}^{\I}$) & (l) OLD ($\gamma_{12t}^{\I}$)\\
\end{tabular}}
\end{center}
\caption{The estimated univariate functional components and the corresponding simultaneous confidence bands in the infection model.}
\label{FIG:CB_covariates}
\end{figure}

Next, we examine the effect of the other predictors in Model (\ref{model:infection}). To test hypotheses $H_0:\gamma_{kt}^{\I}(\cdot)=0$, $k=1,\ldots,12$, we construct the 95\% simultaneous confidence band (SCB) for $\gamma_{kt}^{\I}(\cdot)$'s. In function estimation problems, SCBs are an important tool to quantify and visualize the variability of the functional components; see \cite{Wang:Yang:09}, \cite{Cao:Yang:Todem:12}, and \cite{zheng2016statistical} for some related theory and applications. We consider the period from  03/22/2020 to 04/18/2020 as an illustration and plot the estimated curves for different explanatory variables together with the corresponding SCBs in Figure \ref{FIG:CB_covariates}. Based on Figure \ref{FIG:CB_covariates}, we can observe that at the beginning of the pandemic, the infected cases increase with the population density (PD), which is consistent with our intuition. We also find that the infections increase with African American Ratio and Hispanic Latino Ratio at the beginning of the outbreak.

\begin{figure}[htbp]
\begin{center}
\scalebox{0.85}{
\begin{tabular}{ccc}
\includegraphics[width = 0.35\textwidth]{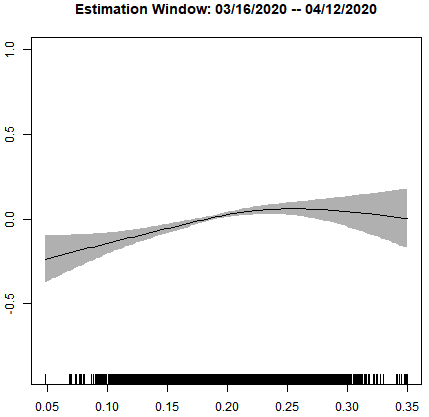} &
\includegraphics[width = 0.35\textwidth]{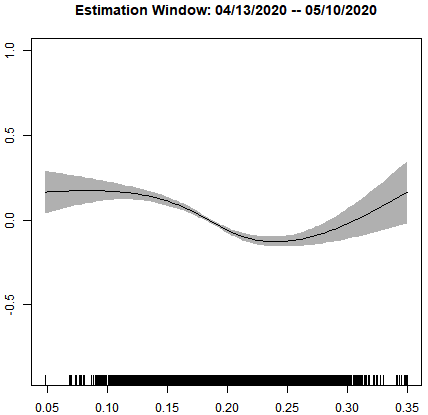} &
\includegraphics[width = 0.35\textwidth]{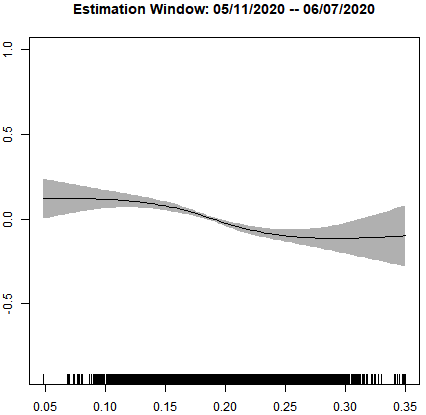}\\
(a) 03/16--04/12  & (b) 04/13-05/10 & (b) 05/11-06/07\\
\includegraphics[width = 0.35\textwidth]{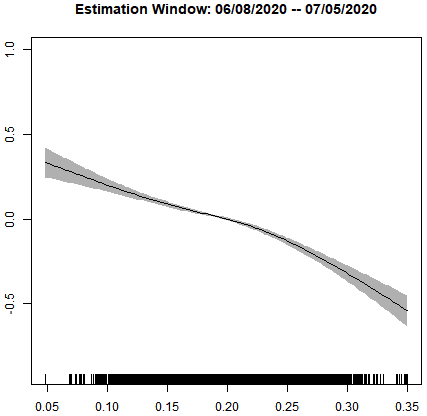} &
\includegraphics[width = 0.35\textwidth]{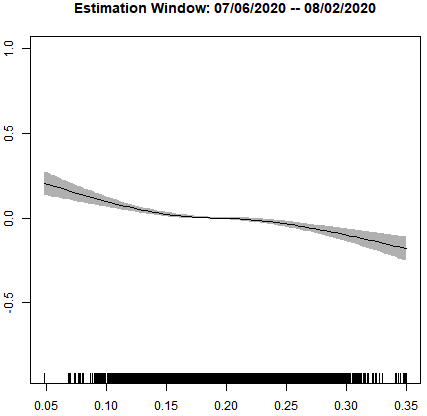} &
\includegraphics[width = 0.35\textwidth]{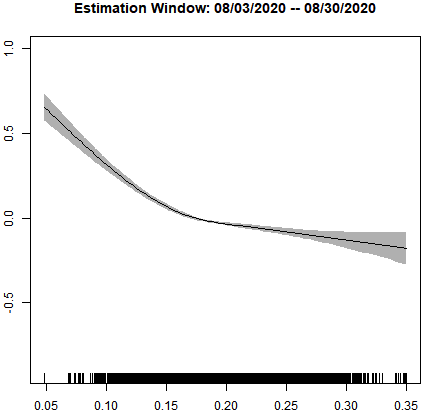}\\
(d) 06/08-07/05 & (e) 07/06-08/02 & (f) 08/03-08/30 ($\gamma_{6t}^{\I}$)\\
\end{tabular}}
\end{center}
\caption{The estimated univariate functional components corresponding to the proportion of the elderly during different periods.}
\label{FIG:CB_OLD}
\end{figure}

We also study the effect of the covariates over time. Figure \ref{FIG:CB_OLD} shows the effect of the aged people rate at different time points over the outbreak. In the early stage of the COVID-19 pandemic, from March to April, COVID-19 struck the elderly more severely than the younger people. 
By mid-April and May, we saw that those communities with less aged people suffered more from COVID-19. Counties with a very high rate of aged people still experience high infection rates. However, when people understood the virus more and took action to protect the senior people, from mid-June to September, those counties with a higher rate of aged people became those least infected.

Movies 1-12 in the Supplementary Material show the estimates and SCBs of the nonparametric functions $\gamma_{kt}^{\I}(\cdot)$, $k=1,\ldots,12$, over the entire study period in the STEM model (\ref{model:infection}).

\begin{figure}[htbp]
\begin{center}
\renewcommand{\arraystretch}{0.3}
\scalebox{0.9}{
\begin{tabular}{ccc}
	\includegraphics[height=1.45in,width=2in]{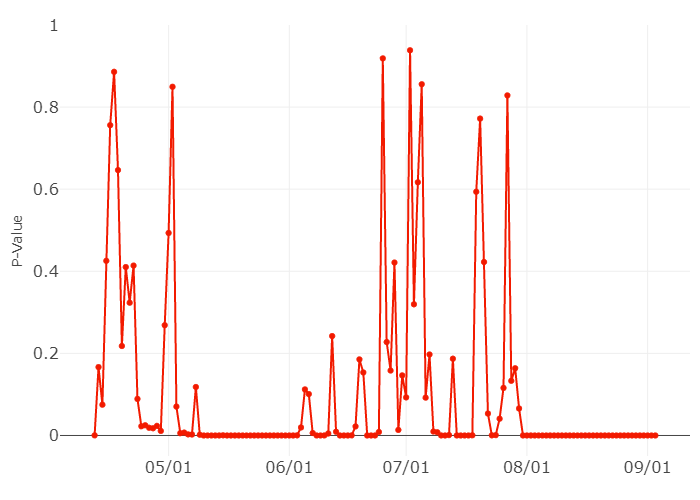} &\includegraphics[height=1.45in,width=2in]{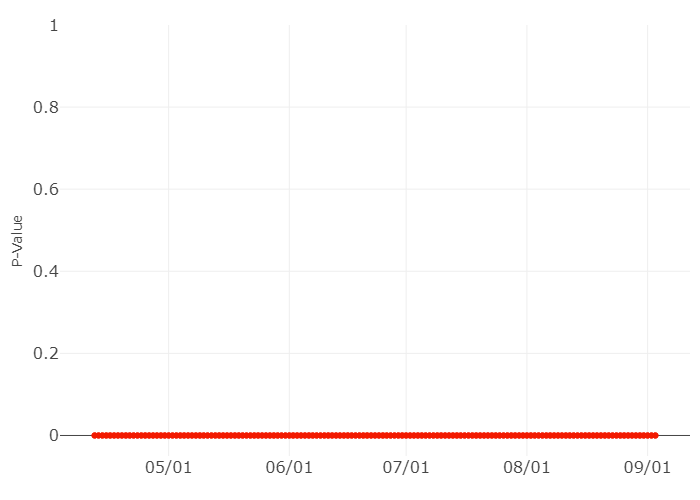} 
	&\includegraphics[height=1.45in,width=2in]{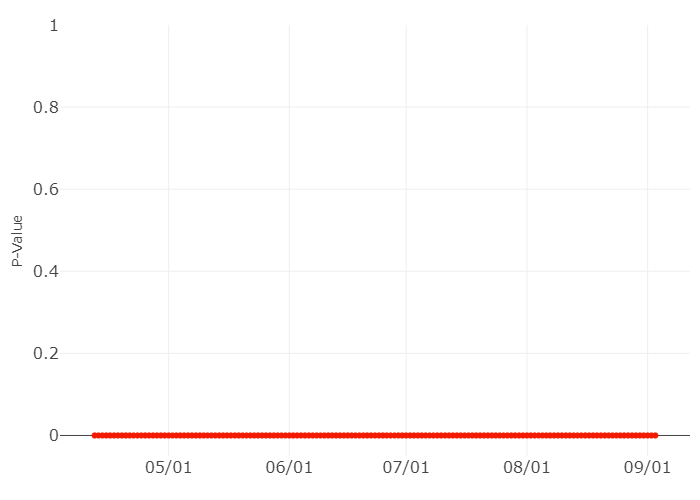}
	\\
	(a) Control ($\alpha_{1t}^{\D}$) & (b) Mobility ($\alpha_{2t}^{\D}$) & (c) Infection ($\beta_{1t}^{\D}$)\\
	\includegraphics[height=1.45in,width=2in]{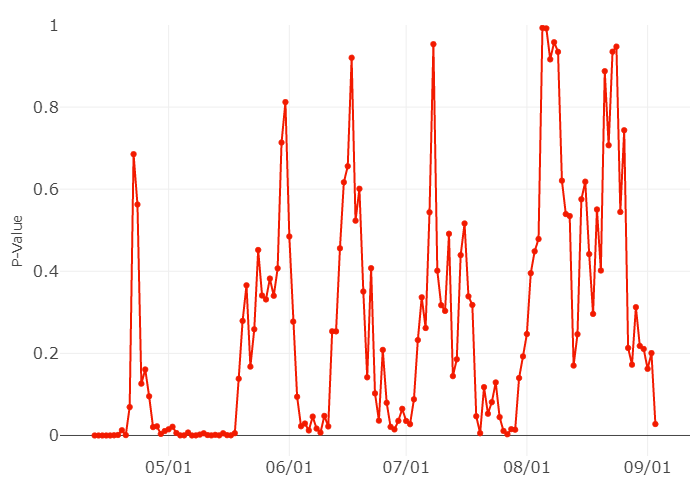}
	& \includegraphics[height=1.45in,width=2in]{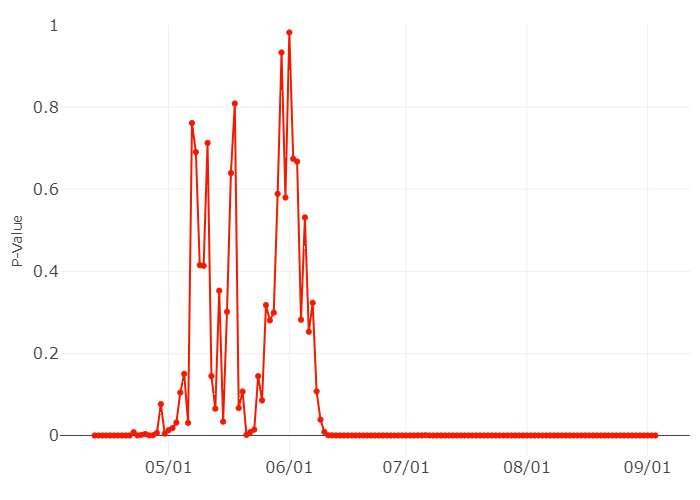} & \includegraphics[height=1.45in,width=2in]{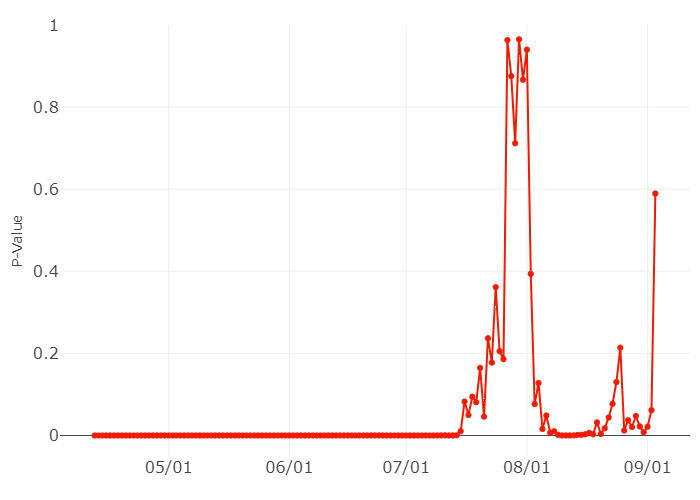}
	\\
	(d) Gini ($\gamma_{1t}^{\D}$) & (e) Affluence ($\gamma_{2t}^{\D}$) & (f) Disadvantage ($\gamma_{3t}^{\D}$)\\
	\includegraphics[height=1.45in,width=2in]{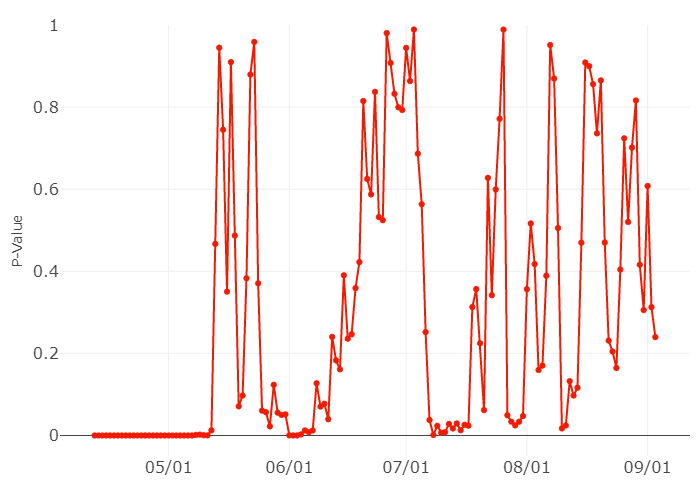}
	& \includegraphics[height=1.45in,width=2in]{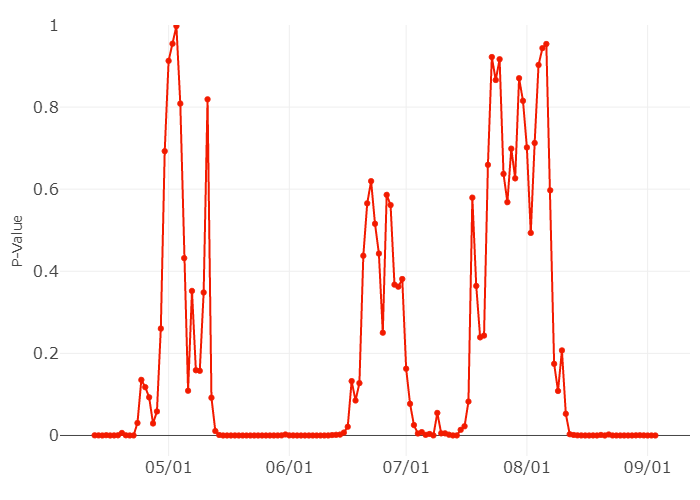} & \includegraphics[height=1.45in,width=2in]{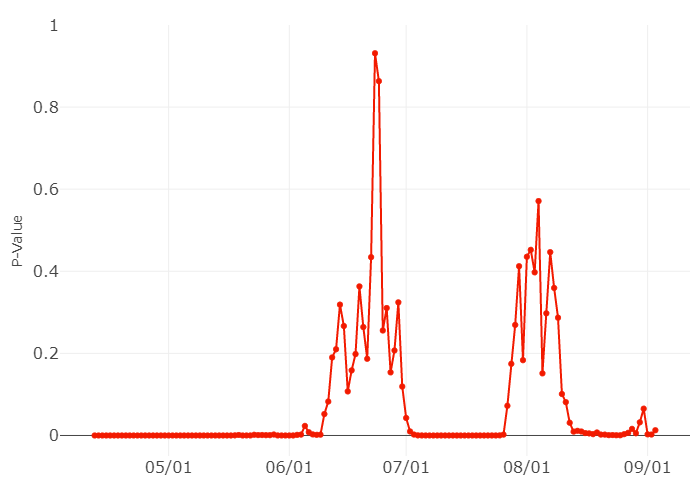}
	\\
	(g) Urban ($\gamma_{4t}^{\D}$) & (h) PD ($\gamma_{5t}^{\D}$) & (i) Tbed ($\gamma_{6t}^{\D}$)\\
	\includegraphics[height=1.45in,width=2in]{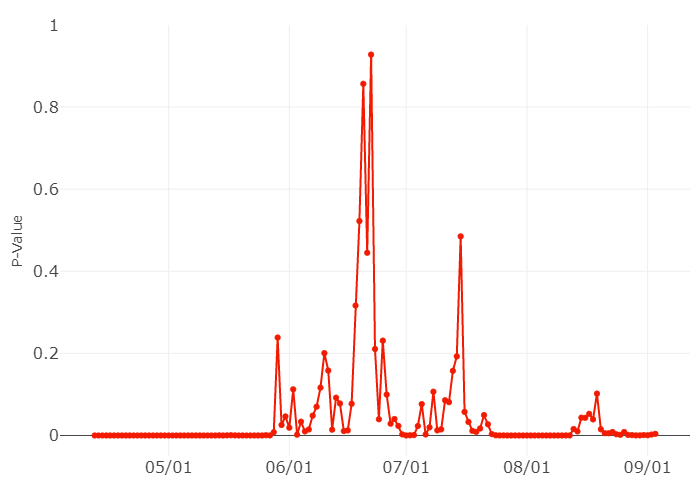}
	& \includegraphics[height=1.45in,width=2in]{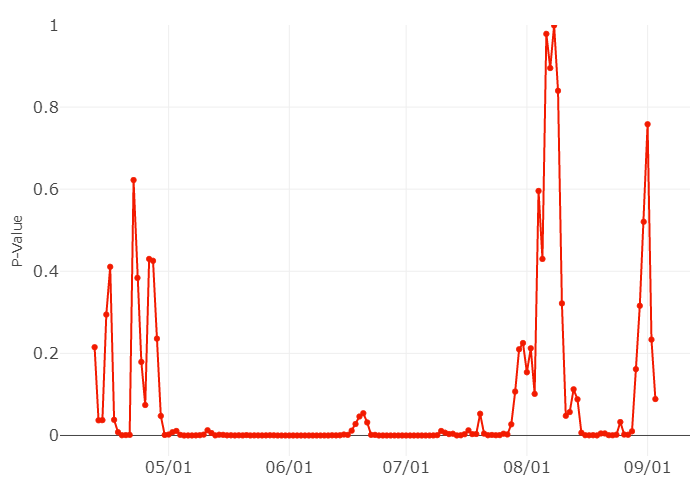} & \includegraphics[height=1.45in,width=2in]{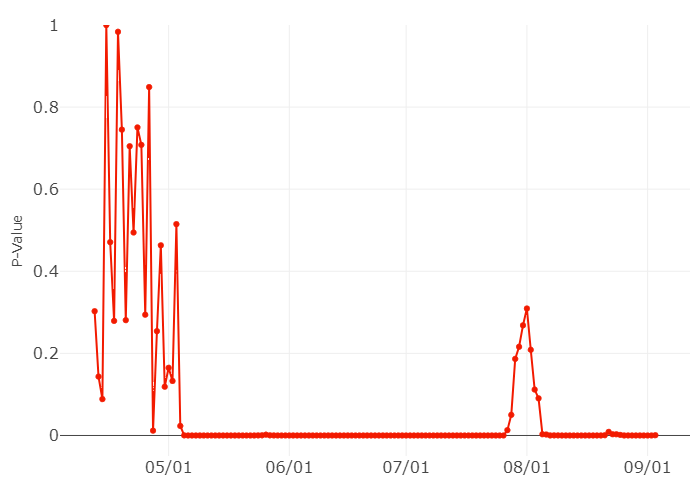}
	\\
	(j) NHIC ($\gamma_{7t}^{\D}$) & (k) EHPC ($\gamma_{8t}^{\D}$)	& (l) AA ($\gamma_{9t}^{\D}$)	\\
	\includegraphics[height=1.45in,width=2in]{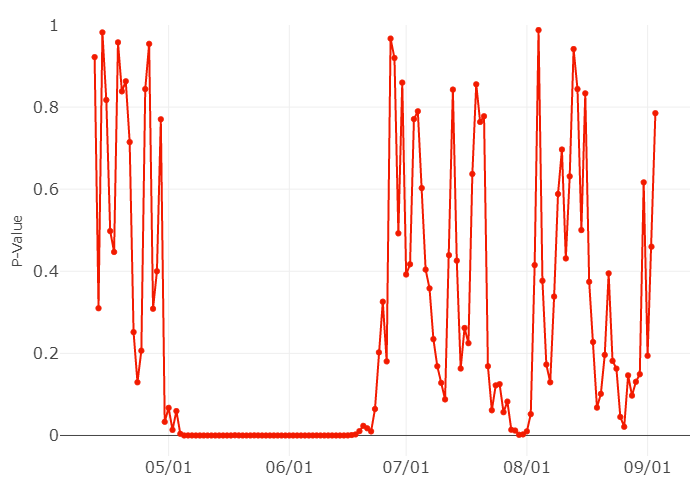}
	&\includegraphics[height=1.45in,width=2in]{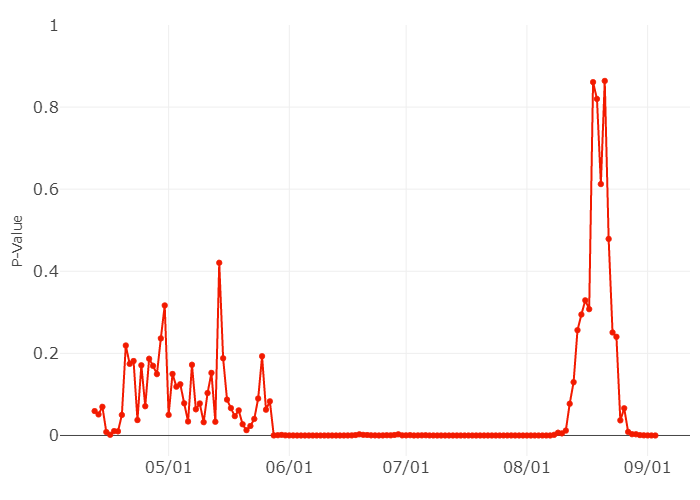} & \includegraphics[height=1.45in,width=2in]{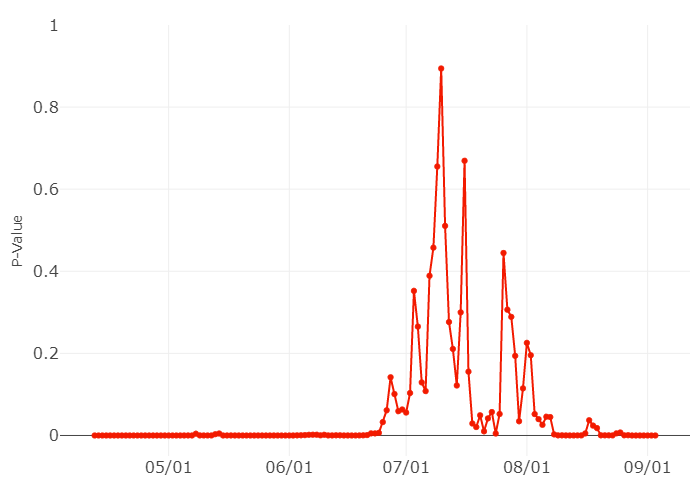}
	\\
	(m) HL ($\gamma_{10t}^{\D}$) & (n) SEX ($\gamma_{11t}^{\D}$) & (o) OLD ($\gamma_{12t}^{\D}$)
\end{tabular}}
\end{center}
\caption{P-values of the hypothesis test of the constant coefficient in model (\ref{model:death}).}
\label{FIG:Death_p_values}
\end{figure}

After the discussion of our finding in the infection model, let us focus on the death model. For Model (\ref{model:death}), we focus on the following hypothesis tests: $H_0: \alpha_{jt}^{\D}=0$, $j=1,2$, $H_0: \beta_{1t}^{\D}=0$ and $H_0: \gamma_{kt}^{\D}=0$, $k=1,\ldots,12$. Figure \ref{FIG:Death_p_values} (b) shows that  ``Mobility" is significant over the entire study period. ``OLD" is significant in the beginning of the pandemic. For other county-level covariates, ``Affluence", ``Disadvantage", ``EHPC", ``AA" and ``SEX" are significant with $p$-values smaller than $0.05$ most of time, while the rest of the predictors are significant on some days, but insignificant on other days.

In addition, movies 13 and 14 in the Supplementary Material illustrate the estimated coefficient functions of $\beta_0^{\I}(\cdot)$ and $\beta_1^{\I}(\cdot)$ in model (\ref{model:infection}). From Movie 13, we can see that the transmission rate, $\beta_0^{\I}(\cdot)$, varies at different locations and in different phases of the outbreak, especially the high rate in late March and April. Movie 14 shows that $\beta_1^{\I}(\cdot)$ also varies from one location to another location, which indicates that the homogeneous mixing assumption of the simple SIR models does not hold. The transmission rate is high in most states at the end of April; however, it became much lower since June. Movie 15 on the Supplementary Material shows the pattern of $\widehat{\beta}_{0t}^{\D}$ in model (\ref{model:death}). From this animation, we observe a severe fatality condition in the southern states in July and a pattern of general decrease in the entire US since August 2020.

\subsection{Short-term Forecasting Performance and Results} 
In this section, we investigate the short-term prediction performance of the proposed method. In the following, we consider $h$-day ahead prediction based on the forecasting method described in Section \ref{sec:prediction}. We choose nine days as a training window for model fit to predict the next seven days, which minimizes the mean squared prediction errors. 

An R shiny app \citep{App_1} is developed to provide a 7-day forecast of COVID-19 infection and death count at both the county level and state level, in which the state level forecast is obtained by aggregating forecasts across counties in each state. This app was launched on 03/27/2020 for displaying the results of our 7-day forecasting.

We demonstrate the accuracy of the STEM for $h$-day ahead predictions, $h=1,\ldots,7$. For comparison, we also consider the two naive models that assume a linear or exponential growth pattern for total confirmed cases for each county: for $i=1,\ldots, n$, 
\begin{itemize}
    \item (Linear) $\mathrm{E}(C_{it}|t) = \beta_{i0} + \beta_{i1}t$,  $\mathrm{Var}(C_{it}|t) = \sigma_{i}^{2}$;
    \item (Exponential, Poisson) $\log\{\mathrm{E}(C_{it}|t)\} = \beta_{i0} + \beta_{i1}t$, $\mathrm{Var}(C_{it}|t) = \exp(\beta_{i0} + \beta_{i1}t)$;
\end{itemize}
and the following simple epidemic method (EM):
\begin{itemize}
    \item (EM) $\log(\mu_{it}) 
	= \beta_0+\beta_1\log(I_{i,t-1})$, 
	$\log(\mu_{it}^{\D}) = \beta_0^{\D}+ \beta_1^{\D}\log(I_{i,t-14})$. 
\end{itemize}

We consider the data collected from April 8 to September 3, 2020. To show the accuracy of different methods, we compute the following root mean-squared prediction errors (RMSPEs):
\[
R_h=T^{-1}\sum_{t=1}^{T}\left\{n^{-1}\sum_{i=1}^{n}(\widehat{Y}_{i,t+h}-Y_{i,t+h})^2\right\}^{1/2}, ~ h=1,\ldots,7,
\]
where $T=119$. 

Table \ref{TAB:RMSPE} shows the average of the RMSPEs for $h$-day. From this table, we can see that our proposed method performs the best in predicting the infection count regardless of the date for prediction. When predicting the death count, both EM and STEM methods outperform the other competitors. Overall, we can see that our proposed method outperforms other competing methods in terms of the prediction accuracy. When we analyze the COVID-19 dataset, we also notice a strong temporal-pattern. To be more specific, at the beginning of this pandemic, a more complex model is necessary to describe the outbreak's different severe levels. As the pandemic expands, a simpler model can be used for future prediction. Start from middle July, a more complex model is necessary due to the pandemic's pattern change. Hence, a more flexible model like the STEM is more appropriate for COVID-19 data analysis.

\renewcommand{\baselinestretch}{1.25}
\begin{table}[h!tb]
\begin{center}
\caption{The average of root mean squared prediction errors ($\text{RMSPE}$) of the infection or death count in the COVID-19 Study, where D$_h$ is for the $h$-day ahead prediction, $h=1,\ldots,7$.}
\label{TAB:RMSPE}
\scalebox{0.85}{
\begin{tabular}{llrrrrrrr}
	\hline \hline
& Method & $\text{D}_1$ & $\text{D}_2$ & $\text{D}_3$ & $\text{D}_4$ & $\text{D}_5$ & $\text{D}_6$ & $\text{D}_7$\\ \hline

\multirow{4}{*}{Infection} 
&Linear &43.844 &59.856 &76.105 &93.301 &111.996 &132.654 &155.276\\
&Exponential &$>$1000 &$>$1000 &$>$1000 &$>$1000 &$>$1000 &$>$1000 &$>$1000\\
&EM &44.270 &81.509 &119.204 &158.006 &197.847 &239.207 &282.382\\
&STEM &35.547 &53.683 &72.200 &91.564 &111.312 &131.843 &154.569\\
\hline
\multirow{4}{*}{Death}
&Linear &2.066 &2.759 &3.418 &4.088 &4.807 &5.563 &6.336 \\
& Exponential &$>$1000 &$>$1000 &$>$1000 &$>$1000 &$>$1000 &$>$1000 &$>$1000 \\
&EM &1.764 &3.053 &4.332 &5.633 &6.972 &8.351 &9.803\\
&STEM &1.411 &2.203 &2.927 &3.615 &4.283 &4.932 &5.629\\
\hline \hline
\end{tabular}}
\end{center} 
\end{table}


\subsection{Long-term Forecasting Performance and Results}

With the rapid spread of the COVID-19 across the US, there has been an increasing public health concern regarding the adequacy of resources to treat infected cases. It is well known that hospital beds, intensive care units, and ventilators are critical for the treatment of patients with severe illness. To predict the timing of the outbreak peak and the number of health resources required at a peak, in this section, we also provide the long-term projection of the death counts, assuming that the future will continue to follow the current pattern, and current interventions will remain the same till the end of forecasting period.

\begin{figure}[htbp]
\begin{center}
\renewcommand{\arraystretch}{0.75}
\begin{tabular}{ccc}
 \includegraphics[width = 0.45\textwidth]{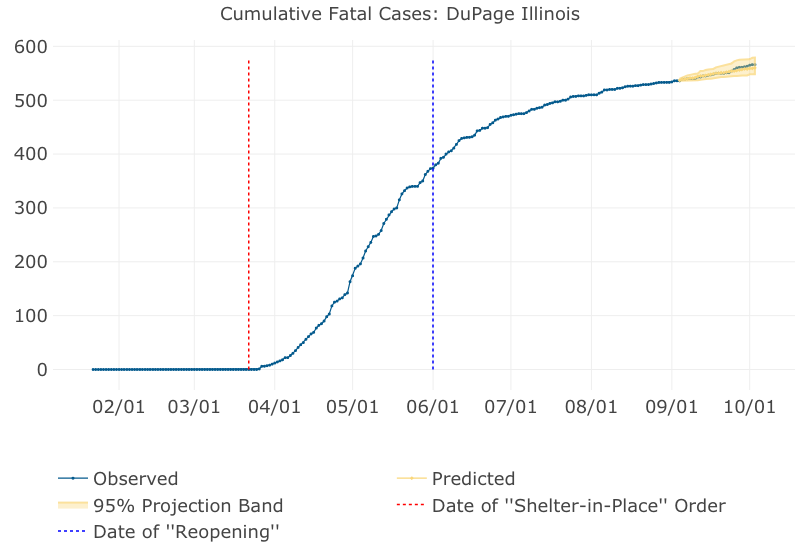} 
	& \includegraphics[width = 0.45\textwidth]{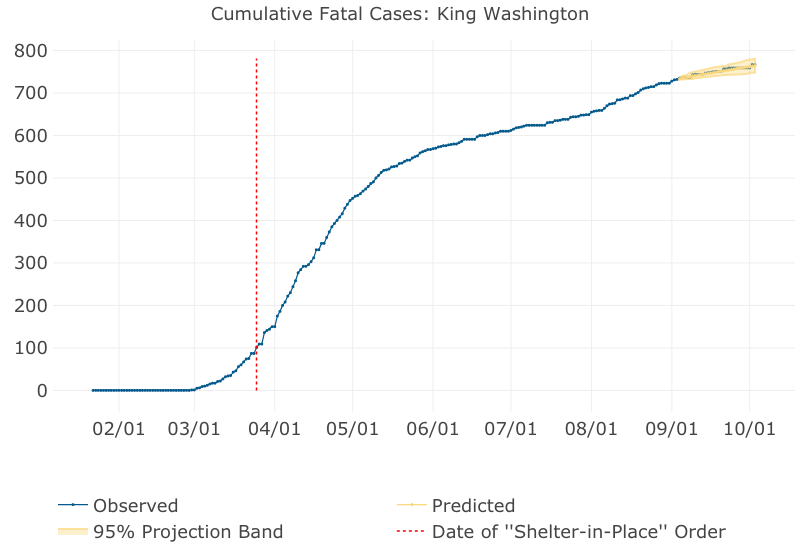} 	\\ 
		(a)  & (b) \\
\includegraphics[width = 0.45\textwidth]{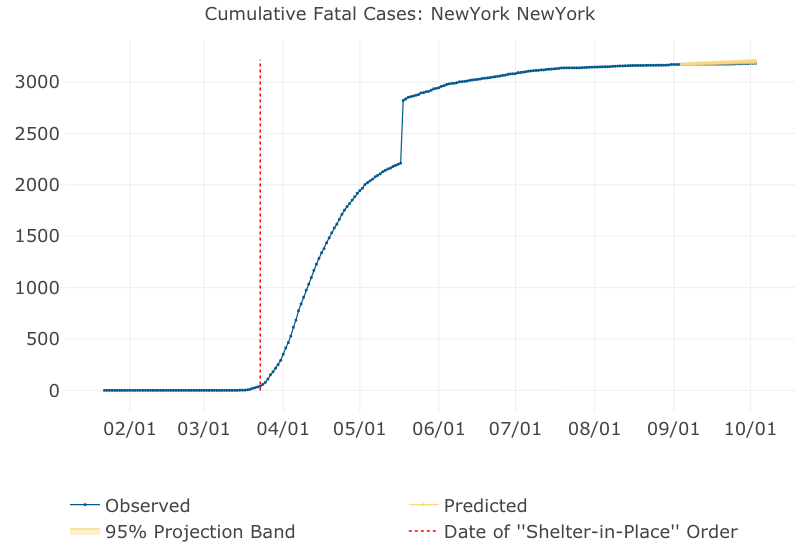}  &\includegraphics[width = 0.45\textwidth]{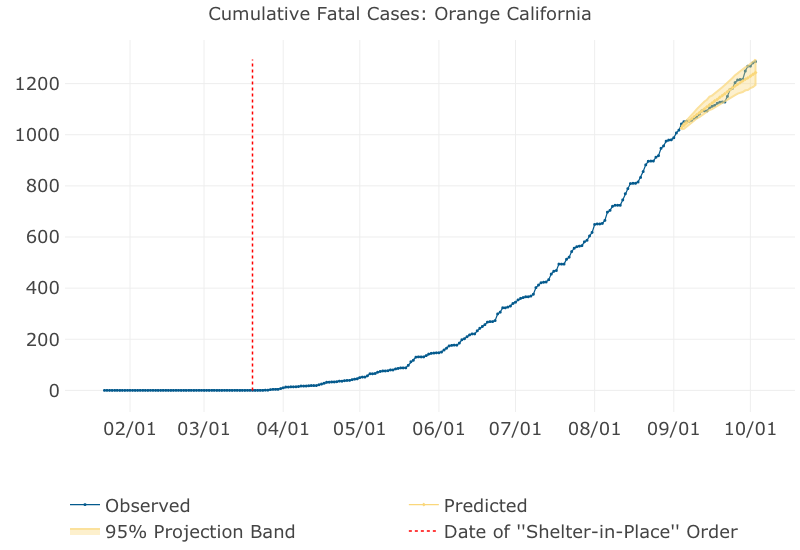} \\
	(c) & (d) \\
	\includegraphics[width = 0.45\textwidth]{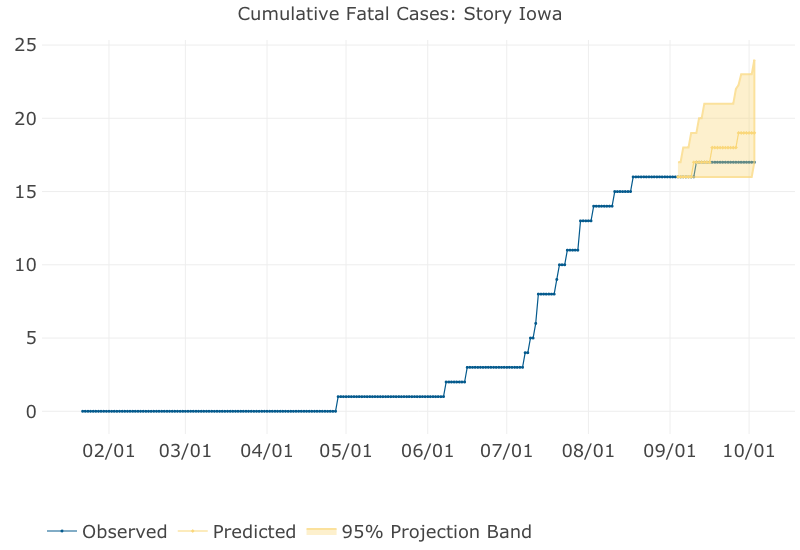} & \includegraphics[width = 0.45\textwidth]{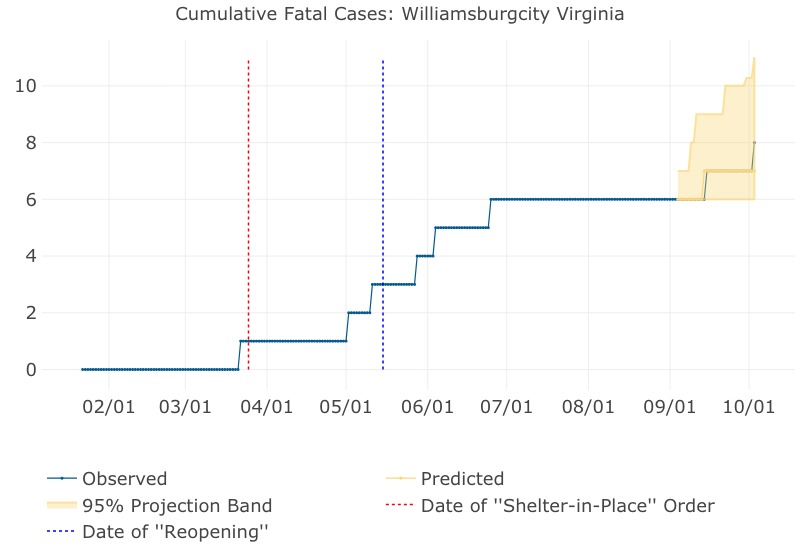}\\
	(e) & (f) \\
\end{tabular}
\end{center}
\caption{Time series plots of cumulative fatal cases with 95\% one month-ahead prediction band in (a) Dupage County, Illinois, (b) King County, Washington, (c) New York City, New York, (d) Orange County, California (e) Story County, Iowa, (f) Willamsburg City, Virginia. The prediction is based on observed data 08/26/2020-09/03/2020.}
\label{FIG:projection_County}
\end{figure}

The long-term projections are conducted at the county and state levels, respectively. Due to the lack of reliable recovered data, we consider the recovery rate uniformly distributed from $5\%\sim15\%$, where the recovery rate is defined as the number of daily recovered cases divided by the number of active cases 14 days (one incubation period) ago. Figure \ref{FIG:projection_County} (a) - (f) show the one month-ahead prediction and the corresponding 95\% prediction band in six different counties. 
Table \ref{Prediction_RMSPE} summarizes the performance of the one month-ahead projection for county-level fatal cases. We use the root of mean squared prediction error to measure the prediction performance. Table \ref{Prediction_RMSPE} shows that our proposed method has good prediction performance. The prediction error increases as time goes by, as the error will accumulate over time. Also, we evaluate the performance of the prediction band. We calculate the empirical coverage rate: the proportion of counties for which the $h$-day ahead prediction band, $h = 1, \ldots, 30$, covers the observed counts. From Table \ref{Prediction_RMSPE}, we see that the empirical coverage rates of the county-level band are close to the nominal level for one week-ahead prediction, and we still have a good coverage rate even for one month-ahead prediction. The projection for other counties and states can be found from our weekly updated Shiny App \citep{App_2} and the webpage of CDC \citep{CDC:20}. Note that \cite{CDC:20} presents predicted death counts for the next four weeks in the US provided by different research teams, including ours, through which we can compare our long-term projection with the other methods. It shows that our approach gives reliable guidance of the trend of death in the future.

\renewcommand{\baselinestretch}{1.25}
\begin{table}[h!tb]
\begin{center}
\caption{Evaluation of one month-ahead prediction.}
\label{Prediction_RMSPE}
\scalebox{0.8}{
\begin{tabular}{cccccccccccc}
	\hline \hline
&  D$_1$ & D$_2$  & D$_3$  & D$_4$  & D$_5$  & D$_6$  & D$_7$  & D$_8$ & D$_9$  & D$_{10}$ \\ \hline
RMSPE & 1.86 & 2.11 & 2.63 & 3.41 & 4.24  & 4.06 & 4.41 & 4.53 & 4.81 & 5.45\\
Coverage
& 0.97& 0.96& 0.95& 0.94&0.94&0.93&0.92&0.91&0.90 &0.90 \\ \hline
 & D$_{11}$  & D$_{12}$ &  D$_{13}$  & D$_{14}$  & D$_{15}$ & D$_{16}$ & D$_{17}$ & D$_{18}$  & D$_{19}$ & D$_{20}$ \\  
\hline
RMSPE & 6.08 &  6.26 & 6.50 & 6.82 & 7.13 & 7.64 & 8.17 & 8.86 & 9.11 & 9.32 \\
Coverage& 0.89&0.88&0.88&0.87&0.87&0.86&0.86&0.85&0.85&0.84\\\hline
& D$_{21}$  & D$_{22}$ & D$_{23}$  & D$_{24}$  & D$_{25}$  & D$_{26}$  & D$_{27}$  & D$_{28}$ & D$_{29}$  & D$_{30}$ \\ \hline
RMSPE & 9.52 & 
9.72 & 9.98 & 10.58 & 11.34 & 11.61 & 11.89 & 12.24 & 12.59 & 13.01 \\
Coverage& 0.84&0.83&0.83&0.82&0.82&0.81&0.80&0.80&0.79& 0.79 \\
\hline \hline
\end{tabular}}
\begin{tablenotes}
\small
\item Note: RMSPE is the daily root of mean squared prediction error for county-level cumulative fatal cases.  Coverage is the empirical coverage rate of estimated $h$-day ahead 95\% projection band for county-level cumulative fatal cases. 
\end{tablenotes}
\end{center} 
\end{table}

\setcounter{chapter}{8} 
\setcounter{section}{7} 
\renewcommand{\thesection}{\arabic{section}} %
\renewcommand{\thesubsection}{8.\arabic{subsection}}%
\setcounter{equation}{0} \renewcommand{\theequation}{8.\arabic{equation}} %
\setcounter{table}{0} \renewcommand{\thetable}{{8.\arabic{table}}} %
\setcounter{figure}{0} \renewcommand{\thefigure}{8.\arabic{figure}} %
\setcounter{algorithm}{0} \renewcommand{\thealgorithm}{8.\arabic{algorithm}} %
\setcounter{theorem}{0} \renewcommand{\thetheorem}{{8.\arabic{theorem}}} %
\setcounter{lemma}{0} \renewcommand{\thelemma}{{8.\arabic{lemma}}} %
\setcounter{proposition}{0} \renewcommand{\theproposition}{{8.\arabic{proposition}}}%
\setcounter{corollary}{0} \renewcommand{\thecorollary}{{8.\arabic{corollary}}}%

\section{Conclusion and Discussion} 
\label{sec:discussion}

This work has aimed to bridge the gap between mathematical models and statistical analysis in the infectious disease study. We created a state-of-art interface between mathematical models and statistical models to understand the dynamic pattern of the spread of contagious diseases. Our proposed model enhances the dynamics of the S(E)IR mechanism through spatiotemporal analysis.

For analyzing the confirmed and death cases of COVID-19, other factors may also be responsible for temporal or spatial patterns. We investigated the spatial associations between the infected count, death count, and factors or characteristics of the counties across the US by modeling the daily infected/fatal cases at the county level in consideration of the county-level factors. Modeling COVID-19 at the county-level and combining local characteristics are very beneficial for the community to understand the dynamics of the disease spread and support decision-making when urgently needed. To examine spatial nonstationarity in the transmission rate of the disease, we proposed a nonparametric spatially varying coefficient model, which allows the transmission to vary from one area to another area. The proposed method can be used as an essential tool for understanding the dynamics of the disease spread, as well as for assessing how this outbreak may unfold through time and space.

Based on our results, disease mapping can easily be implemented to illustrate high-risk areas and thus help policymaking and resource allocation. Our method can also be extended to other situations, including epidemic models in which there are several types of individuals with potentially different area characteristics or more complex models that include features such as latent periods or a more realistic population structure.

Our paper did not take the under-reported issue (for example, the asymptomatic coronavirus infectious cases) into account. Although our model may have partially corrected the problem with the spatiotemporal information, a better way to ultimately solve this problem is to use survey data based on a representative survey sampling strategy \cite{held2019handbook}. For example, following \cite{Finkenstadt:etal:00}, we can assume that the true number of confirmed cases ($C_{it}^{\ast}$) is related to the number of reported cases ($C_{it}$) via $C_{it}^{\ast}=\rho_{it}C_{it}$, where $\rho_{it}$ is the reporting rate, and can be estimated based on the survey data. Then, our STEM can be extended further to this scenario.

\section*{Data Availability Statement}

\begin{itemize}
\item A full list of data citations are available by contacting the corresponding author.  \vskip -.1in

\item The R package ``\texttt{STEM}" of the proposed method can be downloaded from the Github Repository:  \url{https://github.com/covid19-dashboard-us/covid19}. \vskip -.1in

\item The R shiny apps demonstrating the proposed methods can be found from 
\url{https://covid19.stat.iastate.edu/}.
\end{itemize}

\bibliographystyle{asa}
\bibliography{references}

\end{document}